\documentclass[prx,longbibliography,twocolumn,aps,superscriptaddress]{revtex4-1}
\usepackage[utf8]{inputenc}
\usepackage{amsmath}
\usepackage{amssymb}
\usepackage{graphicx}
\usepackage{color}
\usepackage{float}
\usepackage[bookmarks=false,linkcolor=blue,urlcolor=blue,colorlinks,citecolor=blue]{hyperref}
\usepackage{amsmath,mathtools}
\usepackage{dsfont}
\usepackage{amssymb}

\newcommand{\matelem}[3]{\langle #1|#2|#3\rangle}

\newcommand{\average}[1]{\langle #1\rangle}

\newcommand{\nodag}{{\phantom{\dag}}}

\newcommand{\nF}{n_\mathrm{F}}
\newcommand{\nB}{n_\mathrm{B}}

\newcommand{\pois}{\mathrm{pois}}
\newcommand{\qp}{\mathrm{qp}}

\renewcommand{\Re}{\mathop{\text{Re}}\nolimits}

\newcommand{\tr}{\mathop{\text{tr}}\nolimits}

\newcommand{\ket}[1]{\vert {#1}\rangle}
\newcommand{\bra}[1]{\langle{#1}|}

\newcommand{\outprod}[2]{\left\vert {#1} \left\rangle \right\langle {#2} \right\vert}
\newcommand{\inprod}[3]{\left\langle {#1} \left\vert  {#2} \right\vert {#3} \right\rangle}
\newcommand{\abs}[1]{\left\vert {#1} \right\vert}
\newcommand{\av}[1]{\left\langle{#1} \right\rangle}

\newcommand{\Bth}{B_{\text{th}}}
\newcommand{\BQPC}{B_{\text{QPC}}}
\newcommand{\Eq}[1]{Eq.~(\ref{#1})}
\newcommand{\Eqs}[1]{Eqs.~(\ref{#1})}

\newcommand{\eq}[1]{(\ref{#1})}
\newcommand{\Fig}[1]{Fig.~\ref{#1}}

\newcommand{\Sec}[1]{Sec.~\ref{#1}}

\newcommand{\nbrack}[1]{\left(#1\right)}

\newcommand{\nFa}{n_{\text{F},\ell} }
\newcommand{\nFap}{n_{\text{F},\ell'} }
\newcommand\bwt         {\begin{widetext}}
\newcommand\ewt         {\end{widetext}}

\begin{document}
\title{Parity-to-charge conversion in Majorana qubit readout}

\author{Morten I. K. Munk}
\author{Jens Schulenborg}
\affiliation{Center for Quantum Devices, Niels Bohr Institute, University of Copenhagen, 2100 Copenhagen, Denmark}
\author{Reinhold Egger}
\affiliation{Institut für Theoretische Physik, Heinrich-Heine-Universität, 40225 Düsseldorf, Germany}
\author{Karsten Flensberg}
\affiliation{Center for Quantum Devices, Niels Bohr Institute, University of Copenhagen, 2100 Copenhagen, Denmark}
\date{April 5, 2020}

\begin{abstract}
We study the time-dependent effect of Markovian readout processes
on Majorana qubits whose parity degrees of freedom are converted into the charge of a tunnel-coupled quantum dot.
By applying a recently established effective Lindbladian approximation~\cite{Kirsanskas2018Jan,Mozgunov2020Feb,Nathan2020Apr},
we obtain a completely positive and trace preserving Lindblad master equation for the combined dot-qubit dynamics, describing relaxation and decoherence processes beyond the rotating-wave approximation. This approach is applicable to a wide range of weakly coupled environments representing experimentally relevant readout devices. We study in detail the case of thermal decay in the presence of a generic Ohmic bosonic bath, in particular for potential fluctuations in an electromagnetic circuit. In addition, we consider the nonequilibrium measurement environment for a parity readout using a quantum point contact capacitively coupled to the dot charge.
\end{abstract}
\maketitle

\section{Introduction}\label{sec1}

In the pursuit of reliable and scalable qubits, Majorana bound states (MBSs) have received a substantial amount of attention in the previous decade~\cite{Nayak2008Sep,Wilczek2009Sep,Franz2010Mar,Stern2010Mar,Leijnse2012Nov}. Using zero-energy Majorana states, non-abelian many-body braiding statistics could be implemented~\cite{Read2000Apr,Ivanov2001Jan,Kitaev2003Jan,Alicea2011Feb,Flensberg2011Mar,Sau2011Sep,vanHeck2012Mar,Aasen2016Aug}, and quantum information may be encoded in nonlocal degrees of freedom which are robust to local noise~\cite{Kitaev2003Jan,Nayak2008Sep,Terhal2012Jun,Sarma2015Oct,Aasen2016Aug,Landau2016Feb,Plugge2016Nov,Plugge2017Jan,Karzig2017Jun,Litinski2017Sep}.
Several physical Majorana platforms have been proposed and studied over the years~\cite{Kitaev2003Jan,Fu2008Mar,Lutchyn2010Aug,Oreg2010Oct,Cook2011Nov,Choy2011Nov,Vazifeh2013Nov,Klinovaja2013Nov,Beenakker2013Mar,Nadj-Perge2014Oct,Elliott2015Feb,Lutchyn2018May,Fornieri2019Apr}.
Experiments aiming to verify the presence of MBSs have so far focused primarily on measuring zero-bias conductance peaks~\cite{Mourik2012May,Deng2012Dec,Liu2012Dec,Higginbotham2015Sep,Albrecht2016Mar,Deng2016Dec,Nichele2017Sep,Zhang2018Mar} and the fractional Josephson effect~\cite{Rokhinson2012Sep,Wiedenmann2016Jan,Laroche2019Jan}. These phenomena represent key physical effects of zero-energy Majorana end states in one-dimensional (1D) topological superconductors~\cite{Kitaev2001,Kitaev2003Jan,Kwon2004Feb,Lutchyn2010Aug,Oreg2010Oct,Stefanski2016Oct}.
However, despite providing necessary indicators, and with the benefit of readily being experimentally accessible even in the coherent transport regime~\cite{Whiticar2019Feb}, neither zero-bias peaks nor unconventional Josephson relations have so far provided conclusive evidence for the presence of  MBSs~\cite{Akhmerov2011Jan,Das2012Nov,Lee2013Dec,Cayao2015Jan,San-Jose2016Feb,Liu2017Aug,Liu2018Jun,Hell2018Apr,Chiu2019Jan,Vuik2019Nov,Schulenborg2020Jan}.
The ultimate goal thus remains to demonstrate non-abelian braiding statistics, see also Ref.~\cite{Manousakis2020Mar}.

\begin{figure}
\includegraphics[width=\linewidth]{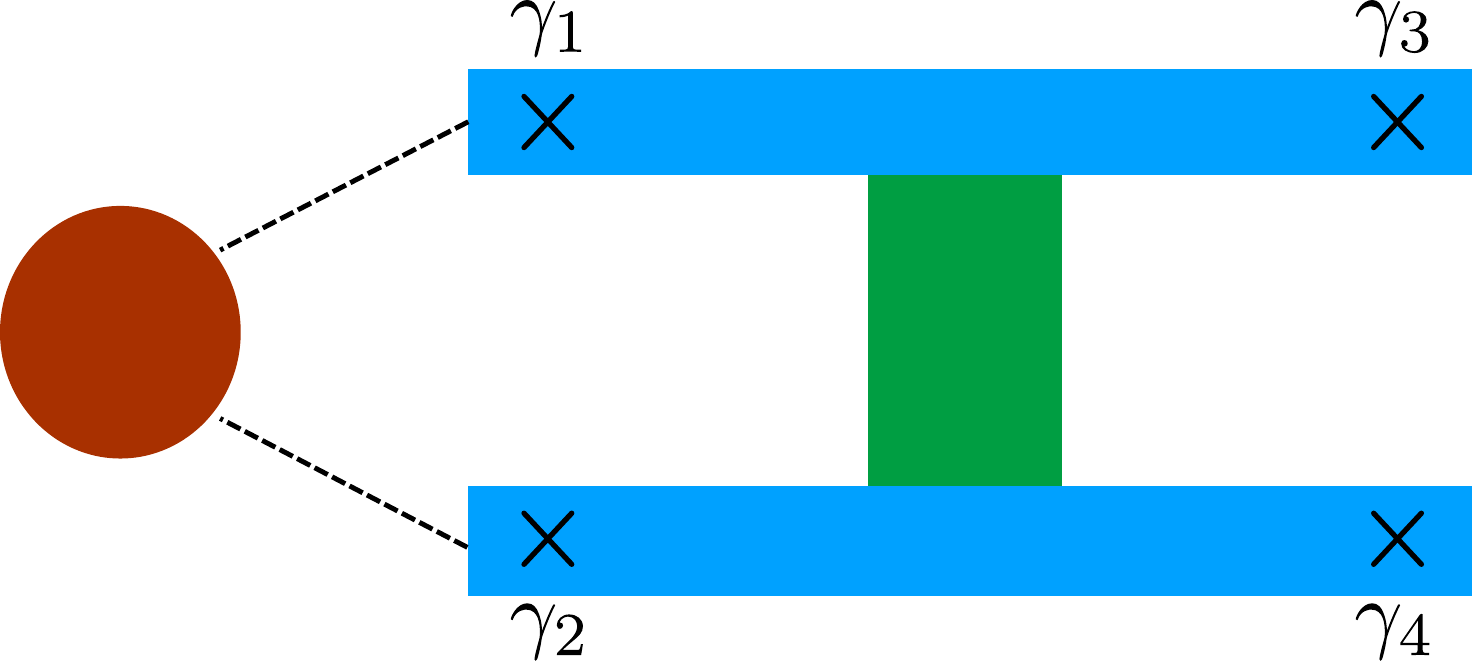}
\caption{Schematic of a Majorana box qubit (MBQ), consisting of two topologically superconducting nanowires (shown in blue), hosting Majorona zero-energy states ($\gamma_i$) at their ends. The nanowires are strongly coupled to a common superconducting ground (green) that effectively provides a common charging energy for the island. The ground-state degeneracy is split by a tunnel-coupled single-level quantum dot (red). }\label{fig1}
\end{figure}

The crux of the latter problem may be solved by developing a reliable readout procedure for the fermion parity of a MBS pair. Indeed, whereas  braiding Majoranas locally in space is very challenging from an experimental perspective, see also Refs.~\cite{Alicea2011Feb,Bauer2018Jul}, alternative schemes have been proposed which simulate braiding purely through parity measurements~\cite{Bonderson2008Jun,Vijay2016Dec,Karzig2017Jun,Knapp2020Mar}. In order to read out the parity of a MBS pair, however, parity has to be converted to a physically observable quantity, such as flux, charge, or capacitance \cite{Szechenyi2019Sep}. This paper focuses on the perhaps simplest Majorana qubit, called the Majorana box qubit (MBQ) \cite{Plugge2017Jan,Karzig2017Jun}, see \Fig{fig1}.
The two-fold degenerate ground state of the MBQ is spanned by the parities of MBS pairs in a system where one has four MBSs with constant total parity. As depicted in \Fig{fig1} and detailed in \Sec{sec2}, one can read out the parity of a MBS pair for any initially prepared qubit state by tunnel-coupling a quantum dot to the respective two MBSs on the island, since this parity in general will affect the outcome of a dot charge measurement~\cite{Flensberg2011Mar,Plugge2017Jan,Karzig2017Jun,Manousakis2017Apr}.

However, a successful readout crucially relies on the total parity in the combined dot-MBQ system being constant over
a sufficiently long measurement time. This means that (i) the readout device itself should not exchange particles with the dot-MBQ system, and (ii) the decoherence due to the readout should be fast compared to decoherence caused by external noise sources which \emph{do} affect the total parity.  Previous theoretical studies of measurement-induced decoherence in Majorana qubits \cite{Plugge2017Jan,Karzig2017Jun,Li2018Nov,Qin2019Apr,Munk2019Apr,Mishmash2020Feb} have analyzed related questions but without taking into account the detailed quantum dynamics of the dot and thereby, in particular, neglecting quantum backaction effects \cite{Clerk2010Apr}.
In this paper, we propose and study a flexible and powerful theoretical approach which can ultimately provide a unified and quite realistic description of the parity-to-charge
conversion process and the corresponding readout dynamics in such a topologically protected system.

 In the main sections~\ref{sec2} and \ref{sec3} of this work, we discuss how a solely capacitively coupled environment representing the readout device causes a prepared quantum state of the dot-MBQ system with fixed total parity to decohere in time.
To describe the decay dynamics, in Sec.~\ref{sec2}, we derive a Markovian quantum master equation~\cite{Bloch1946Oct,Redfield1965Jan,Gorini1976May,Lindblad1976Jun,Breuer2002,Gardiner2004} for the reduced density operator, $\rho(t)$, describing the dot and the two coupled Majorana states. To achieve this, we employ a recently established~\cite{Kirsanskas2018Jan,Mozgunov2020Feb,Nathan2020Apr} effective Lindbladian approximation.
Unlike the common secular approximation~\cite{Breuer2002,Gardiner2004}, this approximation retains nontrivial effects due to the coupling of coherence and population dynamics, i.e., off-diagonal elements of $\rho$ in the local energy eigenbasis couple to diagonal elements, see also Ref.~\cite{Kleinherbers2020Mar}. Moreover, unlike, e.g., the Wangsness-Bloch-Redfield approximation~\cite{Bloch1946Oct,Wangsness1953Feb,Redfield1965Jan}, this scheme is guaranteed to yield completely positive trace-preserving (CPTP) dynamics~\cite{Gorini1976May,Lindblad1976Jun} --- an essential requirement for a physical, probabilistic interpretation of the reduced density matrix $\rho(t)$.
As we are particularly interested in decoherence, i.e., the decay of off-diagonal elements of $\rho(t)$, this approximation is particularly well suited here.
In contrast to previous works \cite{Plugge2017Jan,Karzig2017Jun,Li2018Nov,Qin2019Apr,Munk2019Apr,Mishmash2020Feb}, our approach is able to capture quantum backaction effects on the MBQ state since the quantum dynamics of the dot fermion is taken into account.

Experimentally relevant estimates for relaxation and decoherence rates in this dot-MBS system will be derived for two different types of measurement environments in Sec.~\ref{sec3}. The first is a thermal bath of bosonic modes~\cite{Caldeira1983Sep,Weiss2011Nov}, for which we discuss electromagnetic potential fluctuations in an electric circuit as a concrete example. This case also accounts for the fact that even if the measurement process does not provide the experimentalist with any information, the mere coupling to the measurement device already leads to decoherence due to the inevitable noise in the measurement apparatus.
Second, as an example for a nonequilibrium environment that does provide information about the system state, we consider a voltage-biased,
capacitively coupled quantum point contact (QPC) acting as a dot-charge sensor~\cite{Field1993Mar,Elzerman2003Apr,Ihn2009Sep,Bauerle2018Apr}.
In the outlook section~\ref{sec4}, we then lay out how future work can extend this analysis to a quantitative description, taking into account also other relaxation mechanisms. Such mechanisms could possibly involve particle exchange such as quasiparticle poisoning.  The paper closes with some concluding remarks in Sec.~\ref{sec5}. Finally, we note that technical details have been delegated to two appendices.

\section{Key concepts of MBQ readout}
\label{sec2}

\subsection{Model}\label{sec2a}

The MBQ device of interest is depicted in Fig.~\ref{fig1}. It consists of two topological superconductor nanowires~\cite{Lutchyn2010Aug,Oreg2010Oct}, hosting altogether four zero-energy MBSs at their ends, and an effectively spinless, single-level quantum dot tunnel-coupled to two of the nanowire ends.
With the superconducting bridge, the two nanowires form a single floating island subject to Coulomb charging effects.
We assume that on the  island, the superconducting gap is so large that the influence of quasiparticles and subgap (Andreev) states beyond  MBSs can be neglected.
The low-energy Hamiltonian then reads
\begin{align}
H &= H_0 + H_B + H_I ,\label{HamilDef}\\
H_0 &=   \epsilon\, n_d + \sum_{i=1,2} \gamma_i\left(\lambda_id -\lambda_i^* d^\dagger\right) ,\label{Hamil0}\\
H_I &= \sqrt{g}n_d \varphi.\label{Hamil1}
\end{align}
The Hamiltonian $H_0$ describes both the local coherent dynamics of the dot, with level position $\epsilon$, occupation number operator $n_d=d^\dagger d$,
and fermionic annihilation operator $d$, and of the two tunnel-coupled MBSs.
The latter are described by Majorana operators, $\gamma_i=\gamma_i^\dagger$, with anticommutation relations $\{\gamma_i,\gamma_j\}=2\delta_{ij}$.
The amplitudes $\lambda_{i=1,2}$ for tunneling between dot and $\gamma_i$ are, without loss of generality, parametrized by the real-valued quantities $\lambda$, $a$, and $\phi$,
\begin{align}
\lambda_{1} &= \lambda \geq0, \qquad \lambda_2 = a \lambda e^{i\phi}, \qquad 0\leq a\leq 1.\label{lambdaparam}
\end{align}
Importantly, the phase difference $\phi$ is controllable by, e.g., a variable magnetic flux inside the loop constituted by the tunneling links and the superconducting backbone.
As we show in \Sec{sec2b}, one can tune this phase to split the energies of the MBQ in such a way that it is possible to read out
the MBQ state. For this to work, however, we furthermore require that the charging energy of the superconducting island is large enough to constrain the total fermion  parity of the MBQ,
$(-\mathds{1})^{n_L + n_R}$, where $n_{L/R} = f_{L/R}^\dagger f_{L/R}^{\phantom{\dagger}}$ denotes the occupation of the left/right fermionic state with $f_{L/R} = (\gamma_{1/3} + i \gamma_{2/4})/2$.
We assume Coulomb valley conditions such that all other charge states of the superconducting island in Fig.~\ref{fig1} cost a large excitation energy at least of the order of the charging energy of the island \cite{Plugge2017Jan,Karzig2017Jun}.

The term $H_B$ in \Eq{HamilDef} describes the environment, e.g., representing a measurement device, and $H_I$ is a capacitive coupling between the dot charge and the environment. Concrete implementations of $H_B$ and the specific degrees of freedom, $\varphi$, coupling to the dot charge via $H_I$ are discussed in \Sec{sec2c}. In general terms, the dimensionless coupling constant $g$ in $H_I$ is
determined by the ratio between a capacitive interaction energy, $E_{{\rm int}}$, and a model-specific reference energy, $E_{\text{ref}}$. We require that this reference energy is large in comparison, $0 < E_{\rm int}/E_{\text{ref}} \ll 1$, such that $g \ll 1$ quantifies a weak system-environment coupling, justifying a perturbative expansion in $H_I$.
Furthermore, we here only consider environments $H_B$ which are quadratic in field operators, and that are effectively bosonic from the point of view of the fermions in the dot-MBQ system. As detailed in \Sec{sec2c}, this case includes, for example, photons in a thermal electromagnetic environment as well as the effective bosonic modes originating from the Coulomb interaction between the dot charge and the local electronic charge density in a fermionic environment.

\subsection{Readout principle and fidelity}\label{sec2b}

To explain the readout principle for the MBQ state in concrete terms, in the following we always consider the sector with even total parity, where the total parity of the superconducting island is assumed to be conserved during the entire measurement. In this case, the MBQ has two basis states, $\ket{0_L0_R}$ with $n_L = n_R = 0$ and $\ket{1_L 1_R}$ with $n_L = n_R = 1$.
Given this total-parity constraint, the readout principle and its fidelity rely mostly on the fact that fermionic parity is exchanged through the tunnel couplings between the dot and the tunnel-coupled left Majorana pair, corresponding to $\gamma_1$ and $\gamma_2$ in Fig.~\ref{fig1}. Importantly, the full Hamiltonian \eq{HamilDef} conserves the joined parity of the dot and these two MBSs,
\begin{equation} \label{subparity}
s=(-\mathds{1})^{n_d+n_L}.
\end{equation}
We next observe that due to the presence of  the phase $\phi$ in the tunneling amplitudes \eq{lambdaparam}, eigenstates of the dot-MBQ system Hamiltonian $H_0$ (defined in the absence of the environment) in general have different energies for $s = +1$ and $s = -1$.
Denoting the eigenstates of $H_0$ by $|p,s\rangle$, with $p=\pm$, we have
\begin{align}
H_0|p,s\rangle&= E_{p,s}|p,s\rangle,\quad E_{\pm,s}= \left(\epsilon \pm E_s\right)/2, \notag\\
E_{s=\pm} &= \sqrt{\epsilon^2 +4\lambda^2 (1+a^2+ sa  \sin\phi)}.\label{eq_different_energies}
\end{align}
Since $s$ is a conserved quantity, an initial MBQ state with $s = +1$ will dynamically relax towards a stationary state with $s = +1$ when coupled to the measurement device. By \Eq{eq_different_energies}, this stationary state has an energy different from the energy of the state to which an initial state with $s = -1$ relaxes. In addition, also the average dot occupation number,
\begin{equation}\label{eq_dot_charge}
\av{n_d} = \frac{d \av{H}}{d\epsilon},
\end{equation}
 depends on this energy difference in the long-time limit. This fact ultimately enables one to read out the parity number $s=\pm 1$
 via measurements of the dot charge, see Eq.~\eq{eq_dot_charge}, or via its quantum-capacitive effect $\sim d^2 \av{H}/d\epsilon^2$, see Ref.~\cite{Karzig2017Jun}.

\begin{figure}
\includegraphics[width=\linewidth]{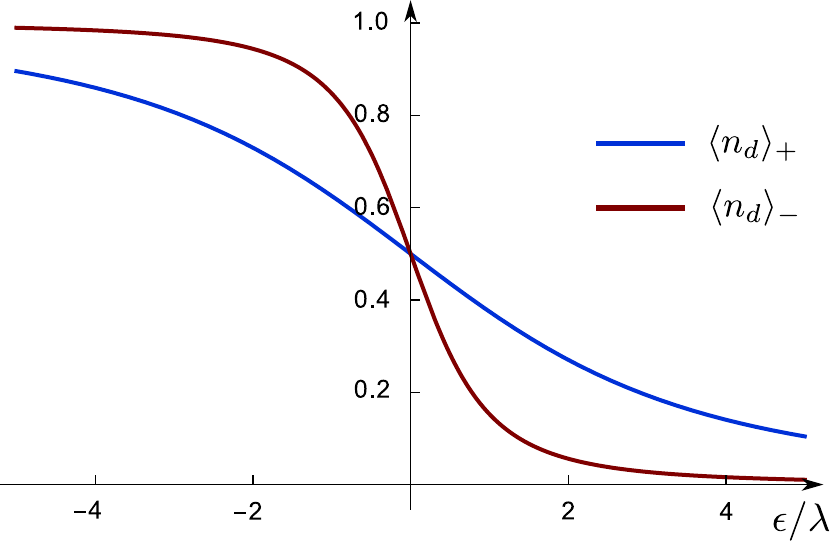}
\caption{Readout principles for the MBQ device in Fig.~\ref{fig1}: Average  steady-state dot occupation number, $\av{n_d}_{s}$, vs $\epsilon/\lambda$ for different parities $s=\pm 1$ in Eq.~\eqref{subparity}.
Here $\av{n_d}_s$ has been calculated for the ground state with energy $E_{-,s}$ in Eq.~\eqref{eq_different_energies}, using $a = 1$ and $\phi = \pi/3$. Calculating  $\av{n_d}_s$ instead for the thermal states $\rho_{\text{st},s}$ in Eq.~\eqref{Thermalstate}, the two curves approach the constant curve $\av{n_d}=1/2$ with increasing temperature. Thereby the readout visibility, i.e., the ability to distinguish the values $s=\pm 1$, will be gradually lost. }\label{fig2}
\end{figure}

Suppose now that the dot is initially empty, $n_d=0$, and the MBQ has been prepared in the initial state
\begin{equation}\label{MBQstate}
\ket{\psi_0}=\alpha_0 \ket{0_L0_R}+\beta_0 \ket{1_L1_R},
\end{equation}
with complex-valued coefficients $\alpha_0$ and $\beta_0$ subject to $|\alpha_0|^2+|\beta_0|^2=1$.
In general, the initial state of the combined dot-MBQ system thus corresponds to a superposition of states with different values of $s=\pm 1$.
Since the energies \eqref{eq_different_energies} of the system depend on $s$, decoherence due to the coupled bath representing the measurement device should relax the reduced density matrix of the dot-MBQ system, $\rho(t)$,  to the stationary limit according to
\begin{align}
\rho(t) = &\begin{pmatrix} \abs{\alpha_0}^2 \rho_{s=+} & \alpha_0\beta_0^*\rho_c\\ \alpha_0^*\beta_0 \rho_c^\dagger & \abs{\beta_0}^2\rho_{s=-} \end{pmatrix}\nonumber\\
\xrightarrow{t\rightarrow \infty} \quad &\rho_{\text{st}}=\begin{pmatrix}\abs{\alpha_0}^2 \rho_{\text{st},s=+} & 0\\ 0 & \abs{\beta_0}^2\rho_{\text{st},s=-} \end{pmatrix}.\label{thermalization}
\end{align}
The diagonal blocks here describe the density matrix projected to the respective subspace with parity $s=\pm 1$, while the off-diagonal part $\rho_c$ describes coherences between
both parity sectors.
The steady-state distributions, $\rho_{\text{st},+}$ and $\rho_{\text{st},-}$, may in practice be distinguished by measuring $\av{n_d}$ or $d\av{n_d}/d\epsilon$, averaged over some time interval. In this way, one performs a projective measurement of the initial MBQ state, where $s=+1$ ($s=-1$) occurs with probability $\abs{\alpha_0}^2$ $\left(\abs{\beta_0}^2\right)$. Once the dot is effectively decoupled from the MBSs by adiabatically adjusting $\epsilon$ towards the limit of zero occupation $n_d = 0$, one knows that the MBQ state equals $\ket{0_L0_R}$ ($\ket{1_L1_R}$) if $s=+1$ ($s=-1$) has been measured.

To better understand the fidelity and limitations of this readout, Fig.~\ref{fig2} shows the dependence of $\av{n_d}_s$ on the dot level energy $\epsilon$, as determined by \Eq{eq_dot_charge} for the ground states corresponding to the energies $E_{-,s}$. We observe that for $\epsilon/\lambda \neq 0$, the ground states in the $s=\pm 1$ sectors
\emph{can} be distinguished by measuring the charge on the dot. If the system is instead prepared in a thermal state,
\begin{equation}\label{Thermalstate}
\rho_{\text{st},s} = \frac{1}{Z_s}\sum_{p=\pm} e^{-\beta E_{p,s}}\outprod{p,s}{p,s},
\end{equation}
with Eq.~\eqref{eq_different_energies} and $\beta=(k_B T)^{-1}$,
the curves in Fig.~\ref{fig2} would flatten towards $\av{n_d}_s=1/2$ as temperature is increased.
Evidently, a charge readout of the dot can still measure $s=\pm 1$ provided that the system has thermalized at a sufficiently low temperature.

We note that for a more general environment, the long-term limit need not be represented by a thermal distribution. Nevertheless, as long as the dot charge $\sim d\av{H}/d\epsilon$ and/or, depending on the setup, its quantum capacitance $\sim d^2 \av{H}/d\epsilon^2 $, differ for $s=+1$ and $s=-1$, the value of $s$ may in principle still be distinguished if the system decoheres to a block-diagonal state as in Eq.~\eqref{thermalization}.

By developing a Lindbladian master equation for the above model, we show in \Sec{sec2e} below that block-diagonal relaxation similar to Eq.~\eqref{thermalization} does indeed generically happen. The missing ingredient for arriving at this  master equation is --- as covered in Sec.~\ref{sec2c} below --- a physical specification of the environment, $H_B$, and its coupling to the dot, $H_I$.
A key advantage of the jump operator approximation established in Refs.~\cite{Kirsanskas2018Jan,Mozgunov2020Feb,Nathan2020Apr}, and summarized in \Sec{sec2e},
is that we may simply write down the master equation once we have determined the environmental correlation function,
\begin{equation} \label{eq-correlator}
	B(t) = \av{\varphi(t)\varphi}.
\end{equation}
This correlator is defined with respect to the initial state of the bath before the measurement begins,
where $\varphi(t) = e^{iH_B t}\varphi e^{-iH_B t}$ is taken in the interaction picture. We note that this procedure directly works only for a vanishing linear moment, $\av{\varphi(t)} = 0$. For $\av{\varphi(t)} \neq 0$, the linear moment needs to be time-independent, $\av{\varphi(t)} = \av{\varphi}$. In that case, one can remove the linear moment, $\varphi\rightarrow \varphi-\av{\varphi}$, by a shift of the dot energy, $\epsilon \rightarrow \epsilon + \av{\varphi}$, in Eq.~\eq{HamilDef}.
As this shift does not introduce an explicit time dependence, the effective Lindbladian approximation in Sec.~\ref{sec2e} will still apply upon using  the bath correlator
\begin{equation}\label{autocor}
	B(t) = \av{[\varphi(t) - \av{\varphi}][\varphi - \av{\varphi}]}
\end{equation}
instead of Eq.~\eqref{eq-correlator}.

\subsection{Physical realizations of environments}\label{sec2c}

\begin{figure}
\includegraphics[width=\linewidth]{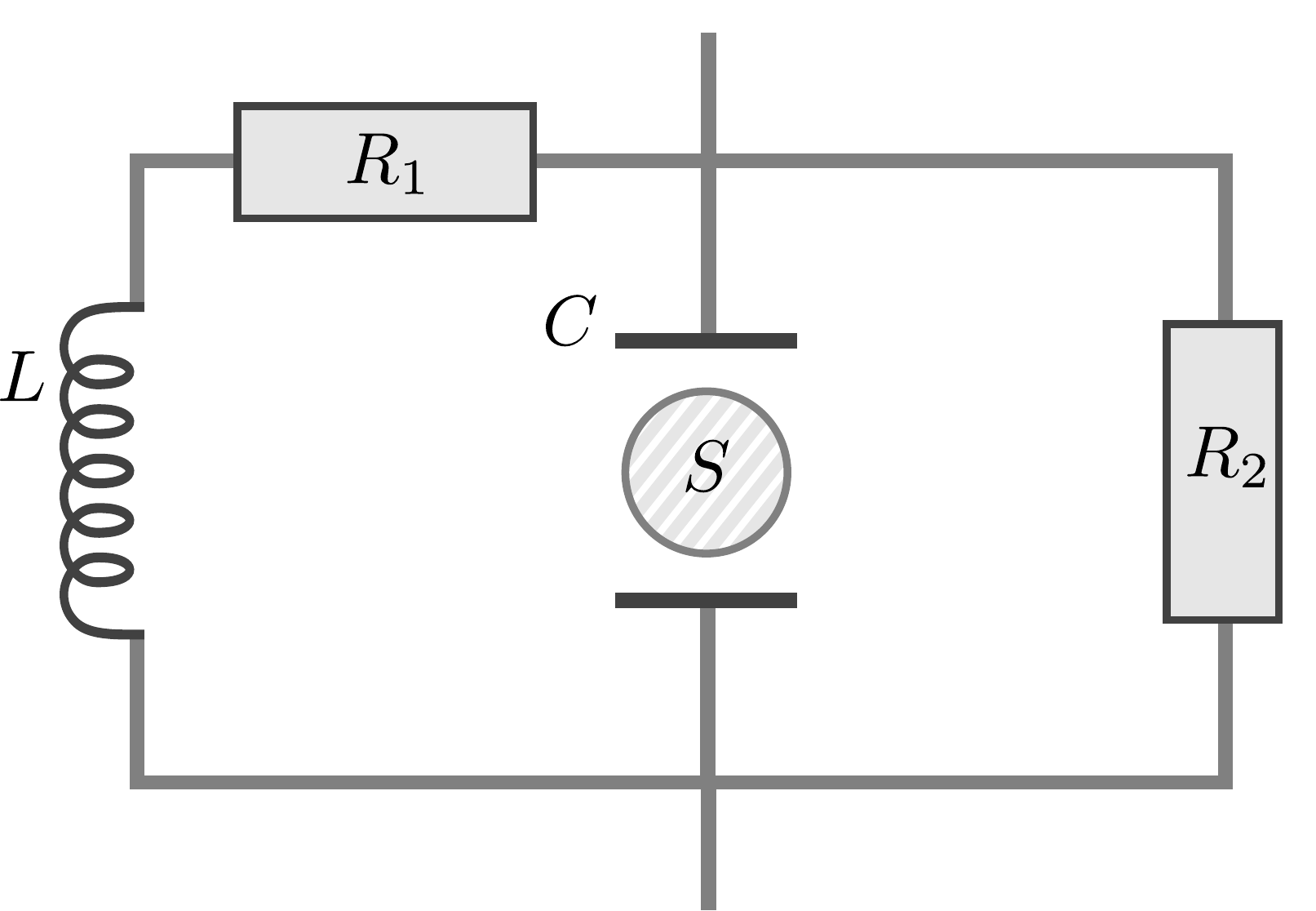}
\caption{Schematic circuit representing a typical electromagnetic environment. The dot-MBQ system ($S$) is capacitively coupled to the environmental inductance and resistances.
The resulting potential $\varphi$ on the capacitor $C$ enters as a charge coupling in the Hamiltonian \eqref{HIEM}.  }\label{fig3}
\end{figure}

Let us now precisely formulate the physical systems representing the readout device. As announced in Sec.~\ref{sec1}, we consider two different cases. The first is an Ohmic thermal bath of bosonic modes. This can be seen as a simple phenomenological model for the effects of a measuring apparatus on the dot-MBQ system, such as capacitive noise due to voltage fluctuations in the electronic circuit coupled to the dot, see Fig.~\ref{fig3}. A bosonic bath can, however, also be taken at face value, as a microscopic model of thermal relaxation of the system which will invariably be present due to charge couplings with the environment.
The second addressed case is that of a nonequilibrium measurement environment, formed by two voltage-biased electronic leads coupled by a QPC. Since the QPC is also capacitively coupled to the dot, see \Fig{fig4}, the QPC transmission is affected by the Coulomb interaction with the dot.  By monitoring the conductance through the QPC, one can thereby measures the dot charge and hence the parity $s=\pm 1$.

\subsubsection{Thermal bath of bosons}\label{sec2c1}

Let us first consider a bosonic environment in thermal equilibrium. This model is useful both for describing the inherent decoherence due to electromagnetic radiation, but also as a simple phenomenological model for understanding the dynamics of the system under a generic readout.
The environmental Hamiltonian,
\begin{equation}
	H_B = \sum_q \omega_q\left(b^\dagger_q b_q^{} + \frac{1}{2}\right),\label{eq_hamiltonian_thermal_bath}
\end{equation}
in this case consists of non-interacting bosons characterized by quantum numbers $q$ and energies $\omega_q$,  where $b^\dagger_q$ and $b^{}_q$ are the corresponding creation and annihilation operators. We assume that the dot charge capacitively couples to these bosons,
\begin{equation}\label{eq_hamiltonian_thermal_interaction}
	H_I = \sqrt{g}n_d \varphi ,\quad \varphi = \sum_q \left( M_q^{} b^\dagger_q + M^*_q b^{}_q\right).
\end{equation}
The bath operator $\varphi$ is determined by the mode-dependent coupling energies $M_q$. The small dimensionless coupling constant $g$, which we have introduced in \Eq{HamilDef}, is physically related to the ratio of the capacitive interaction energy, $E_{\rm int}$, of the dot-environment coupling and the typical frequency $\omega_0$ of the environmental oscillators, $g = g(E_{\rm int}/\omega_0)$.
The spectral density associated with the coupling is assumed to be Ohmic with some cutoff function $\mathcal{C}(\omega,\omega_c)$ determined by a cutoff frequency $\omega_c$ (where $\omega_c\approx \omega_0$),
\begin{equation}
	J(\omega) = \sum_q |M_q|^2\delta(\omega - \omega_q) = \omega \mathcal{C}(\omega,\omega_c).\label{eq_thermal_spectral_density}
\end{equation}
An Ohmic spectral density occurs in many different scenarios~\cite{Caldeira1983Sep,Weiss2011Nov}, including, e.g., the capacitive coupling of the system to an electromagnetic transmission line~\cite{Gardiner2004}. The precise form of the cutoff function $\mathcal{C}$ depends on the physical nature of the bath, as we further discuss below. At this stage, it is only relevant in so far as it will regularize integrals at high frequencies in what follows.

The initial density operator of the bath, $\rho_B$, taken before the dot couples to the environment at times $t \ge 0$, is assumed to be thermal, $\rho_B = e^{-\beta H_B}/\mathrm{Tr}\left[e^{-\beta H_B}\right]$.
The expectation value $\langle\varphi(t_1)\dotsc\varphi(t_N)\rangle$ with respect to $\rho_B$ thus disappears by virtue of Wick's theorem for any odd number $N$ of bath operators, with $\varphi(t) = e^{iH_B t}\varphi e^{-iH_B t}$. As stated above and detailed in \Sec{sec2e}, the relaxation of the dot-MBQ system then only depends on the auto-correlation function,
\begin{align}
	&\Bth(t) = \left\langle\varphi(t)\varphi(0)\right\rangle\label{eq_thermal_bath_correlation}\\
	&= \int_0^\infty d\nu J(\nu)\left[e^{i\nu t}\nB(\nu) + e^{-i\nu t}(\nB(\nu) + 1)\right],\nonumber
\end{align}
with the index ``th'' indicating the case of a thermal bath.
Equation \eq{eq_thermal_bath_correlation} derives from the vanishing two-point correlators $\langle b^\dagger_qb^\dagger_q\rangle = \langle b_qb_q\rangle = 0$,  and the occupations $\langle b^\dagger_q b_q\rangle = \langle b_qb^\dagger_q\rangle - 1 = \nB(\omega_q)$, with the Bose-Einstein distribution,
$\nB(\omega) = (e^{\beta\omega} - 1)^{-1}$.
The relaxation of the dot-MBQ system is then determined by the Fourier transform of Eq.~\eqref{eq_thermal_bath_correlation},
\begin{align}
	\Bth(\omega) &= \int_{-\infty}^{\infty}dt\, \Bth(t)e^{i\omega t} \notag\\
	&= \frac{\pi }{2}\omega \mathcal{C}(\omega,\omega_c)\left[\coth\nbrack{\frac{\omega}{2k_BT}} + 1\right].\label{thermalBomega}
\end{align}

To obtain physically meaningful estimates for $g$ and for the cutoff function $\mathcal{C}$, we next observe that in many situations of practical interest, the dominant bosonic reservoir is represented by the electromagnetic modes in the electric circuit connected to the dot-MBQ system.
In such cases, the specific form of $g$ and  $J(\omega)$ may often be derived from the electrodynamical properties of the equivalent classical circuit. As a specific example, consider the case sketched in Fig.~\ref{fig3}, where the dot-MBQ system is placed in an $LC$ circuit and couples through the voltage drop $\varphi/e$ over the capacitor $C$ to the bath. The interaction Hamiltonian may then be written as
\begin{equation}\label{HIEM}
H_I =  n_d  \varphi.
\end{equation}
In thermal equilibrium, the correlation function $B(t) = \av{\varphi(t)\varphi}$ may be calculated by using the Kubo formula and the fluctuation-dissipation theorem.
The impedance of the circuit in Fig.~\ref{fig3} is given by
\begin{equation}
Z(\omega) = \left( \frac{1}{R_2} + \frac{1}{R_1 + i\omega L} + i\omega C\right)^{-1},
\end{equation}
and $B(\omega)$ follows as (see also Ref.~\cite{Weiss2011Nov})
\begin{equation}\label{BLC}
	B_{LC}(\omega) =  e^2\omega\Re\left(\frac{\abs{Z(\omega)}^2}{Z(\omega)}\right) \left[\coth\left(\frac{\omega}{2k_B T}\right) + 1\right].
\end{equation}
Rescaling $\varphi \mapsto \sqrt{g} \varphi$ in Eq.~\eqref{HIEM} and $B_{LC}(\omega) \mapsto B_{LC}(\omega)/g$ in Eq.~\eqref{BLC},
this expression matches the general result for a thermal bosonic bath with Ohmic spectral density in Eq.~\eqref{thermalBomega}, where the coupling constant $g=g_{LC}$ and the bath cutoff frequency $\omega_c$ are given by
\begin{equation}\label{glcwc}
    g_{LC} = \frac{e^2}{2C\omega_{LC}},\quad \omega_c=\omega_{LC}= \frac{1}{\sqrt{LC}},
\end{equation}
and $\omega_{LC}$ is the $LC$ resonance frequency of the circuit in Fig.~\ref{fig3}.  As expected,
the dimensionless small system-bath coupling, $g_{LC}$, follows as
the ratio between the capacitive interaction energy, $E_{\rm int}=e^2/2C$, and the reference energy  set by the $LC$ resonance frequency, $E_{\rm ref}=\omega_{LC}$.
Equation~\eqref{BLC} predicts a cutoff function ${\cal C}(\tilde\omega)$, with $\tilde \omega=\omega/\omega_c$, of the form
\begin{equation}\label{w12}
	\mathcal{C}(\tilde{\omega}) = \frac{4}{\pi} \frac{\tilde{\omega}_2 + \frac{\tilde{\omega}_1}{1+\tilde{\omega}_1^2 \tilde{\omega}^2} }{\left(  \tilde{\omega}_2 + \frac{\tilde{\omega}_1}{1+\tilde{\omega}_1^2 \tilde{\omega}^2} \right)^2 + \tilde{\omega}^2 \left( 1 - \frac{1}{\tilde{\omega}_1^{-2} + \tilde{\omega}^2} \right)^2},
\end{equation}
with $\tilde{\omega}_{i=1,2} = (R_{i}C\omega_{LC})^{-1}$.  Evidently, ${\cal C}(\tilde \omega)$ approaches a constant for $\tilde\omega\to 0$ but decays $\propto 1/\tilde\omega^2$
for $\tilde\omega\to \infty$. This limiting behavior is characteristic for a Lorentzian cutoff function.

\subsubsection{QPC detector}\label{sec2c2}

\begin{figure}[t!!]
\includegraphics[width=\linewidth]{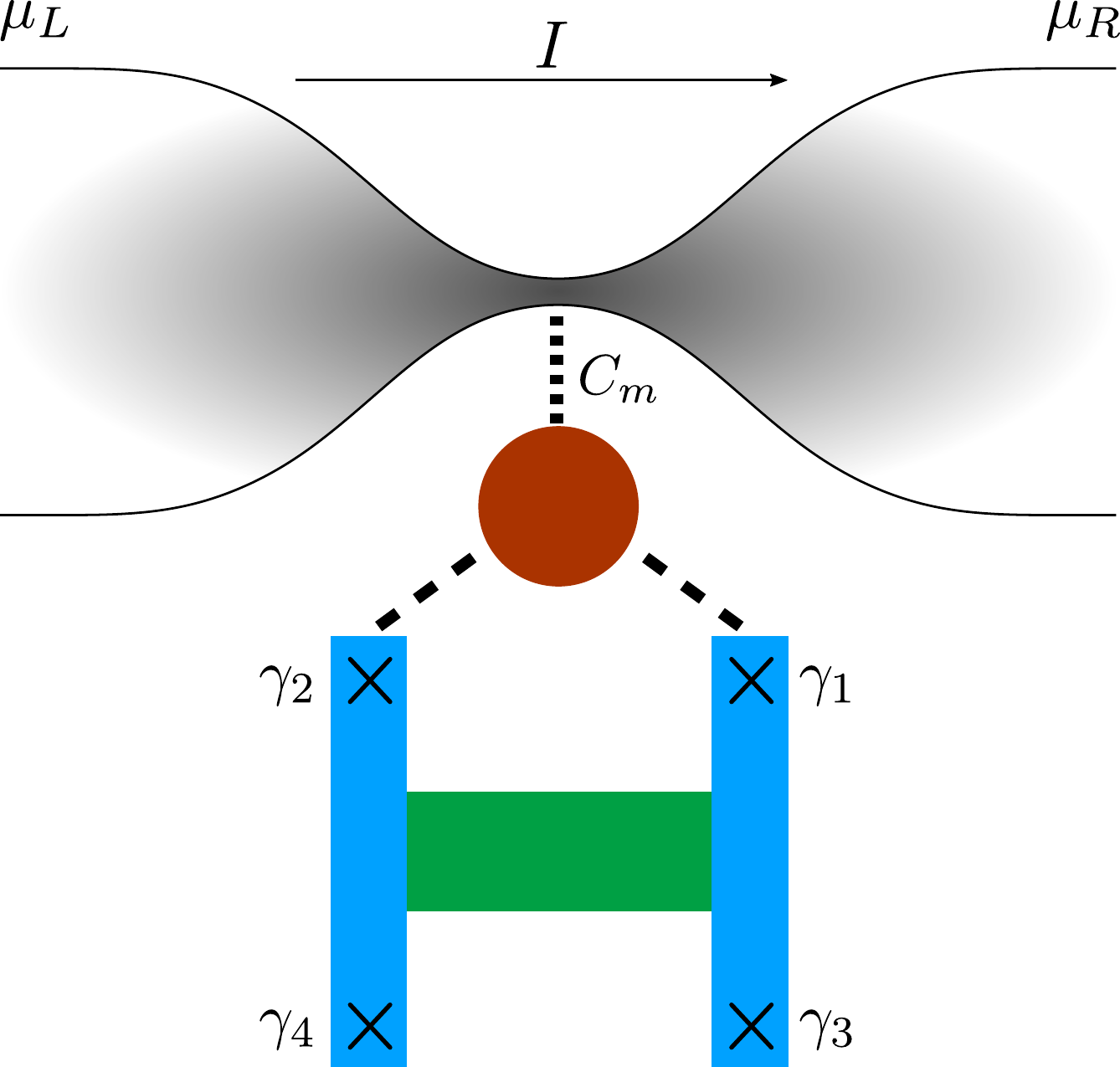}
\caption{Schematic Majorana parity readout using a quantum point contact (QPC) connecting two voltage-biased leads.
The dot-MBQ system capacitively couples to the QPC through a mutual capacitance $C_{m}$ between the charge density in
the QPC and the charge on the dot. This coupling decoheres the dot-MBQ system and perturbs the potential that the QPC feels,
leading to a parity-dependent shift of the conductance through the QPC. }\label{fig4}
\end{figure}

Next we consider a measurement apparatus defined by a QPC that weakly couples together two voltage-biased electronic leads, see \Fig{fig4}.
The QPC is also capacitively coupled to the dot charge.
This coupling mechanism in turn affects the QPC transparency, and hence the measured conductance through the QPC. One can thereby perform a readout of the parity $s=\pm 1$ in Eq.~\eqref{subparity}. As explained above, the outcome of this measurement also determines the eigenvalue of the MBS parity operator $i\gamma_1\gamma_2$.

To good accuracy, the setup in Fig.~\ref{fig4} can be modeled by \cite{Field1993Mar,Elzerman2003Apr,Ihn2009Sep,Bauerle2018Apr}
\begin{gather}
	H_B = \sum_{k;\ell=L,R} \epsilon_{\ell k}c^\dagger_{\ell k}c_{\ell k}^{\phantom{\dagger}}\quad,\quad H_I = \sqrt{g}n_d\varphi,\label{eq_hamiltonian_qpc_interaction}\\
	\quad g = (E_{\rm int}/E_{\rm ref})^2 \quad,\quad \varphi = E_{\text{ref}}V\hat\rho. \notag
\end{gather}
The bath here corresponds to the left and right electronic leads together with their mutual coupling via the QPC.
The Hamiltonian $H_B$ contains the annihilation (creation) operators $c^{(\dagger)}_{\ell k}$ for electrons in single-particle eigenstates of the combined lead-QPC-lead system. The corresponding eigenenergies, $\epsilon_{\ell k}$, are labeled by the wave vector $k$ and the index $\ell = L,R$. This index specifies whether the scattering state originates
from the left or the right lead. The bath operator $\varphi$ in Eq.~\eqref{eq_hamiltonian_qpc_interaction} contains the local electron density operator $\hat \rho$ in the small (essentially point-like) region representing the central QPC region, with volume $V$.  We assume that the capacitive interaction between $\hat \rho$ and the dot charge represents
the dominant coupling between the QPC and the dot-MBQ system, see also App.~\ref{appA}.
The corresponding interaction energy, $E_{\rm int} = 2e^2/(C_m V)$, is  determined by the mutual dot-QPC capacitance $C_m$ per volume $V$, where the factor $2$ accounts for the electron spin.
To justify the weak-coupling approximation, $g \ll 1$, the energy $E_{\rm int}$ must be small compared to a reference energy $E_{\text{ref}}$. The latter energy is  obtained from the following analysis.

We assume that scattering states originating from the left/right lead thermalize according to Fermi-Dirac distributions with equal temperatures, $T_L = T_R = T$, but different chemical potentials, $\Delta\mu = \mu_L - \mu_R \geq 0$. This potential bias induces a stationary charge current across the QPC. The envisioned readout relies on the fact that the capacitive coupling of the QPC to the dot affects the QPC transparency, and hence the current response to the potential bias depends on the dot occupation~\cite{Field1993Mar,Elzerman2003Apr,Ihn2009Sep}.
As detailed in App.~\ref{appA}, the bath correlators describing the time-dependent effect of this readout on the dot-MBQ system are obtained by expressing $\hat\rho$ in terms of the operators $c_{\ell k}^{}$ and $c_{\ell k}^\dagger$. We find a time-independent linear moment, $\langle\varphi(t)\rangle = \langle\varphi\rangle \neq 0$. As pointed out in \Sec{sec2b}, one can absorb $\langle\varphi\rangle$ by a shift of the dot level energy $\epsilon$.
The Fourier transform of the auto-correlation function \eqref{autocor} is then found as
\begin{align}
&\BQPC(\omega) = \int_{-\infty}^\infty dt\, \BQPC(t) e^{i\omega t}\label{eq_qpc_bath_correlation_fourier}\\
&= \pi\sum_{\ell,\ell'=L,R} J_{\ell\ell'}(\omega) \left[\coth\left(\frac{\omega+\mu_{\ell\ell'}}{2k_B T}\right)+1 \right],\notag
\end{align}
with the lead-dependent spectral densities
\begin{align}
	&J_{\ell\ell'}(\omega) = \int_{-\infty}^\infty d\Omega \;\Gamma_{\ell\ell'}\left(\Omega + \mu_{\ell\ell'} - \frac{\omega}{2},\Omega + \mu_{\ell\ell'} + \frac{\omega}{2}\right)\notag\\
	&\times\left[\nF\left(\Omega - \frac{\Delta\mu_{\ell\ell'}+\omega}{2}\right) - \nF\left(\Omega + \frac{\Delta\mu_{\ell\ell'} + \omega}{2}\right) \right],\label{eq_qpc_spectral_density}
\end{align}
where we use the Fermi-Dirac distribution, $\nF(\omega) = (e^{\beta\omega}+1)^{-1}$,
the lead-averaged chemical potentials $\mu_{\ell\ell'} = (\mu_{\ell} + \mu_{\ell'})/2$, and the potential differences $\Delta\mu_{\ell\ell'} = \mu_{\ell} - \mu_{\ell'}$.
The coupling function
\begin{equation}
\Gamma_{\ell\ell'}(\omega,\omega') = E_{\text{ref}}^2\sum_{kk'}|\tau_{\ell k,\ell' k'}|^2\delta(\omega - \epsilon_{\ell k})\delta(\omega' - \epsilon_{\ell'k'})\label{eq_qpc_coupling}
\end{equation}
describes how the QPC scatters electrons from lead $\ell$ with energy $\omega$ into lead $\ell'$ with energy $\omega'$.
In App.~\ref{appA}, we explicitly evaluate Eq.~\eqref{eq_qpc_coupling} for the case of 1D leads with the QPC approximated by a $\delta$-peak potential.
In general, $\Gamma$ scales with the energetic densities of states in the respective lead, $D_\ell=D_\ell(E=\mu_{\ell})$,
and with the typical transmission coefficient $\tau$ of the QPC,
$\Gamma \sim \tau^2 (E_{\text{ref}}D_\ell)(E_{\text{ref}} D_{\ell'})$. A small coupling $g$ can then be realized in two different ways: The first is to have low QPC transparency $\tau\ll 1$, as set by precise implementation of the QPC.  Alternatively, one needs a reference scale $E_{\text{ref}}$ that is small compared to $1/D_{L,R}$ but at the same time large compared to the capacitive energy $E_{\rm int}$, thus leading to $g = (E_{\rm int}/E_{\text{ref}})^2 \ll 1$ according to \Eq{eq_hamiltonian_qpc_interaction}.
Physically, this corresponds to either a relatively low density of states or to a large mutual capacitance $C_mV$.

To understand how $\Gamma_{\ell\ell'}$ in \Eq{eq_qpc_spectral_density} behaves as a function of $\Omega$, and hence how it enters the spectral densities $J_{\ell\ell'}$, we note that the Fermi functions in \Eq{eq_qpc_spectral_density} will effectively restrict the support of the integrand to the window
\begin{equation}\label{supportwindow}
-\frac{\omega + |\Delta\mu_{\ell\ell'}|}{2} < \Omega < \frac{\omega + |\Delta\mu_{\ell\ell'}|}{2}.
\end{equation}
Under the assumption that the applied voltage bias and any internal energy scale determining the QPC transparency (e.g., a potential barrier height) are much smaller than the average chemical potential with respect to the band bottom of the leads, $\Delta\mu_{\ell\ell'} \ll \mu_{\ell\ell'}$,  we can distinguish two limits, namely the cases $|\omega| \ll \mu_{\ell\ell'}$ and $|\omega| \gg \mu_{\ell\ell'}$.
For small frequencies, $|\omega| \ll \mu_{\ell\ell'}$, the coupling profile $\Gamma_{\ell\ell'}$ in \Eq{eq_qpc_spectral_density} can be assumed $\Omega$-independent within
the region \eqref{supportwindow} where the integrand has significant support,
$\Gamma_{\ell\ell'} \sim [E_{\text{ref}}D_\ell(\mu_{\ell\ell'})]^2 \sim (E_{\text{ref}}/\mu_0)^2$, with the average chemical potential $\mu_0 = (\mu_L + \mu_R)/2$.
We here assumed a form of the density of states as appropriate for a 1D electron gas with $|\Delta\mu_{\ell\ell'}|\ll \mu_{\ell\ell'}$,  where one finds
$D_\ell(\mu_{\ell\ell'})\sim 1/\mu_{\ell\ell'}$.
For large frequencies, $|\omega| \gg \mu_{\ell\ell'}$, on the other hand, the coupling factor
$\Gamma_{\ell\ell'}(\Omega + \mu_{\ell\ell'} - \omega/2,\Omega + \mu_{\ell\ell'} + \omega/2)$ in Eq.~\eqref{eq_qpc_spectral_density}
is expected to decay as $1/|\omega|$ for most $\Omega$.  One can rationalize this fact by noting that the density of states decreases, similarly to the case of a 1D Fermi gas, with $1/\sqrt{|\omega|}$ for sufficiently strong lateral electron confinement in the QPC.
Importantly, to regularize Eq.~\eqref{eq_qpc_bath_correlation_fourier} at high frequencies, we also need to account for the finite electronic bandwidth that eventually cuts off the integral.

To qualitatively include all the above-mentioned effects, we now set $E_{\text{ref}} = \mu_0/2$ with $\mu_0=(\mu_L+\mu_R)/2$ and introduce an exponential cutoff.
We thus consider the simplified coupling function
\begin{equation}
	\Gamma_{\ell\ell'}\left(\Omega + \mu_{\ell\ell'} - \frac{\omega}{2},\Omega + \mu_{\ell\ell'} + \frac{\omega}{2}\right) \rightarrow \frac{1}{4}e^{-\frac{|\omega|}{\omega_c}},\label{eq_qpc_coupling_cutoff}
\end{equation}
with  $\omega_c \simeq \mu_0$. The dimensionless coupling constant introduced in \Eq{HamilDef} then equals
\begin{equation}
g = \nbrack{2E_{\rm int}/\mu_0}^2,\label{eq_qpc_coupling_constant}
\end{equation}
where  $E_{\rm int} = 2e^2/(C_mV)$. The weak-coupling assumption holds for $E_{\rm int}\ll \mu_0$.
For a quantitatively more precise calculation, one can resort to a specific QPC model as shown, e.g., in App.~\ref{appA}, followed by a numerical evaluation of \Eq{eq_qpc_spectral_density}.

Here we proceed by inserting \Eq{eq_qpc_coupling_cutoff} into \Eq{eq_qpc_spectral_density}. We then obtain an Ohmic spectral density with a potential shift and an exponential cutoff,
\begin{align}
	J_{\ell\ell'}(\omega) &\rightarrow \frac{1}{4}(\omega + \Delta\mu_{\ell\ell'}) e^{-\frac{|\omega|}{\omega_c}}\notag \\
	&\overset{\Delta\mu_{\ell\ell'}\ll\omega_c}{\approx} \frac{\omega + \Delta\mu_{\ell\ell'}}{4} e^{-\frac{|\omega + \Delta\mu_{\ell\ell'}|}{\omega_c}}.\label{eq_qpc_spectral_density_final}
\end{align}
Using this result in Eq.~\eq{eq_qpc_bath_correlation_fourier}, summing over $\ell,\ell'=L,R$, and comparing the result to
\Eqs{eq_thermal_spectral_density} and \eq{eq_thermal_bath_correlation}, we observe that $\BQPC(\omega)$  becomes a lead average of bosonic bath correlators in thermal equilibrium, $\Bth$ in \Eq{thermalBomega},
\begin{align}
\BQPC(\omega) &= \frac{2\Bth(\omega) + \Bth(\omega + \Delta\mu) + \Bth(\omega - \Delta\mu)}{2}\label{eq_qpc_bath_correlation_fourier_final}
\end{align}
with the potential bias $\Delta\mu = \mu_L - \mu_R$.

In summary, \Eq{eq_qpc_bath_correlation_fourier_final} states that the readout procedure represented by the potential gradient $\Delta\mu$ manifests itself analogously to the capacitive noise of thermal fluctuations. For $\Delta\mu \gg T$, and when lead-state energy differences $-\Delta\mu < \omega < \Delta\mu$ are most relevant for the readout, this contribution to the bath noise and to the relaxation of the dot-MBQ system becomes dominant. In this regime, we show in \Sec{sec3b2} that $\Delta\mu$ plays the role of an effective temperature for the decay rates. In the opposite high-temperature limit, $T \gg \Delta\mu$, the dynamics instead represents a purely thermal decay due to the two leads, $\BQPC\approx 2\Bth$. However, the readout may still work if the dot charge, and hence the QPC conductance, depends on the final MBQ state, and thus on the parity of the initial MBQ state.

\subsection{Mapping to spin-boson model}\label{sec2d}

In this subsection, we  show that our model \eqref{HamilDef} is intimately related to the celebrated spin-boson model, which is a paradigmatic
model for describing the dissipative dynamics of two-level quantum systems \cite{Weiss2011Nov}.
To that end, we first observe that the joined parity $s$ in Eq.~\eqref{subparity} is conserved for the model in Eq.~\eqref{HamilDef},
\begin{equation}
[H,s]=0, \qquad s=(-\mathds{1})^{n_d+n_L}.
\end{equation}
Our system, defined by the dot and the two coupled MBSs, can be described in terms of two different two-level systems which are both coupled to a common bosonic bath. Below, we  make this connection explicit. The dynamical properties of the spin-boson model have been thoroughly studied in the past~\cite{Weiss2011Nov,Egger1994Nov,Grifoni1998Oct,DiVincenzo2005Jan,Lindner2018Sep}. In contrast to those studies, we here encounter
two copies of the spin-boson model, corresponding to the parity eigenvalues $s=\pm1$, respectively.  The dynamics of coherences between those two subsectors then represents the quantity of most interest. Note that such coherences do not violate parity superselection rules~\cite{Wick1952Oct,Aharonov1967Mar,Streater2000Dec} since they comply with total parity conservation once the parity $(-\mathds{1})^{n_R}$ of the uncoupled Majorana pair is accounted for.

Introducing the auxiliary Majorana operators $\eta_{1,2}$ for representing the dot fermion, $d = (\eta_1 + i\eta_2)/2$, we first define the Pauli operator algebra
\begin{align}\label{pauli0}
\tilde{\sigma}_x &= -i\gamma_1 \eta_2,\quad \tilde{\sigma}_y=i \gamma_1 \eta_1 ,\quad \tilde{\sigma}_z = - i\eta_1 \eta_2.
\end{align}
Next we write the parity operator \eqref{subparity} as $s=-\gamma_1\gamma_2\eta_1\eta_2$ in order  to express Eq.~\eqref{HamilDef} as
\begin{align}\label{RotHam}
&H = -\frac{1}{2} (\epsilon +  \sqrt{g}\varphi)\sigma_z  - \frac{\Delta_s}{2} \sigma_x + H_B,\\
&\Delta_{s=\pm 1} = 2\lambda\sqrt{1+a^2+2sa\sin\phi}, \nonumber
\end{align}
where we use the rotated Pauli operators
\begin{eqnarray}\label{pauli2}
\sigma_{\alpha=x,y,z} &=& e^{i\theta_s\tilde \sigma_z} \tilde{\sigma}_\alpha e^{-i\theta_s \tilde\sigma_z},\\  \nonumber
\theta_{s} &=&  - \frac{1}{2} \tan^{-1}\left( \frac{ a  \cos\phi }{s+ a \sin\phi}
\right).
\end{eqnarray}
We note that a constant energy shift has been neglected in Eq.~\eqref{RotHam}, along with the term $\sqrt{g}\varphi/2$. Indeed, upon averaging over the bath degrees of freedom, the  last term yields a contribution $\sim n_d\av{\varphi(t)}$ up to order ${\cal O}((\sqrt{g})^2)$.
Since the average $\av{\varphi(t)}$ is time-independent for all environments considered here, see Sec.~\ref{sec2c},
such a contribution only generates a shift of $\epsilon$ which can be calibrated away.

For $\Delta_+\neq \Delta_-$ in Eq.~\eqref{RotHam}, the system state relaxes to the stationary limit \eqref{thermalization} for standard reasons. In particular, since the energies of the two blocks with $s=\pm 1$ do not match, there are no cancellations of dynamical phases in the off-diagonal entries of the density matrix. The large number of bosonic modes then implies that these terms will cancel out in the long-time limit.
However, for $\Delta_+= \Delta_-$, the evolution of the off-diagonals blocks is identical to the diagonal blocks, and the long-time limit of the density matrix is instead given by
\begin{align}
\rho(t) \quad \xrightarrow[\Delta_+=\Delta_-]{t\rightarrow \infty}\quad  &P_0 \outprod{0_d}{0_d} \outprod{\psi_0}{\psi_0} \nonumber\\
+ &P_1 \outprod{1_d}{1_d} \outprod{\overline{\psi}_0}{\overline{\psi}_0},
\end{align}
where $n_d$ is the occupation of the dot, $P_{0,1}$ the probability to encounter $n_d=0,1$ in the readout,  $|\psi_0\rangle$ has been specified in Eq.~\eqref{MBQstate}, and we use $\ket{\overline{\psi}_0} = \alpha_0 \ket{0_L1_R} + \beta_0 \ket{1_L 0_R}$.
Thus the dot occupation can be read out, but no information will be gained in this case. In fact, the final step of emptying the dot will simply restore
the initial MBQ state.

\subsection{Effective Lindbladian}\label{sec2e}

We are now in a position to derive the quantum master equation governing the time evolution of the reduced density matrix, $\rho(t)$, describing the dot-MBQ system under the influence of the dissipative environment.
In general, a master equation describing a CPTP Markovian time evolution of $\rho(t)$ can always be cast into Lindblad form~\cite{Gorini1976May,Lindblad1976Jun,Breuer2002},
\begin{subequations}\label{LindbladForm}
\begin{align} \rho(t) &= e^{\mathcal{L}t} \rho,\\
\mathcal{L} \rho &= -i[H_{LS} + H_0, \rho]\notag\\
&+ \sum_k\Gamma_k \left( L_k^{} \rho L_k^\dagger-\frac{1}{2}\{L_k^\dagger L^{}_k, \rho\} \right),
\end{align}
\end{subequations}
where the jump operators $L_k$ describe dissipative transitions  induced by the environment.
The corresponding transition rates are
non-negative, $\Gamma_k \geq 0$, thereby guaranteeing CPTP time evolution.
Furthermore,
the Lamb shift contribution appearing in the coherent part of the time evolution is captured by a Hamiltonian $H_{LS}$.
This term encodes system energy renormalizations due to the dressing of system operators by environmental modes.
Such effects may occur even at zero temperature.

 Conventional recipes for deriving Markovian master equations for open quantum systems, such as the Wangsness-Bloch-Redfield approach~\cite{Bloch1946Oct,Wangsness1953Feb,Redfield1965Jan}, in general do not result in master equations of Lindblad form and hence do not necessarily yield CPTP evolution. In contrast, the effective Lindbladian approximation, previously established in Refs.~\cite{Kirsanskas2018Jan,Mozgunov2020Feb} and very recently put on a rigorous footing by Nathan and Rudner~\cite{Nathan2020Apr}, automatically stipulates a Lindbladian form, and thus does away with such problems.
 In this subsection, we give a brief overview of this approximation and apply it to our model.
In effect, the approximation \emph{prescribes} the form of the jump operator,
\begin{align}
L&=  \frac{\sqrt{g}}{2}\sum_{m,n} \sqrt{ B(E_n-E_m)} \inprod{m}{-\sigma_z}{n} \outprod{m}{n},\label{NLJumpOp}
\end{align}
where  $B(\omega) = \int_{-\infty}^\infty dt \, e^{i\omega t} B(t)$. The states $\ket{n}=\ket{p,s}$ are energy eigenstates of the system Hamiltonian $H_0$,
see Eq.~\eqref{eq_different_energies}, and $\sigma_z$ has been defined in Eq.~\eqref{pauli2}, see also App.~\ref{AppB} for a detailed discussion.
The appearance of the square root of the Fourier transformed boson correlator \eqref{autocor}
can be rationalized by noting that Fermi's Golden Rule is then immediately recovered for the transition rates between eigenstate populations.
While jump operators of the form in Eq.~\eqref{NLJumpOp} have been suggested before \cite{Kirsanskas2018Jan}, one of the central contributions of Nathan and Rudner~\cite{Nathan2020Apr} is to put this approximation on solid theoretical grounds by providing an error bound on $\dot{\rho}(t)$.
The approximation consists (i) of a familiar type of Markovian approximation, which is equivalent to the Wangsness-Bloch-Redfield approach in the sense that both approaches share the same error bound $\mathcal{E}_M$. However, Ref.~\cite{Nathan2020Apr} formulates (ii) another approximation that is not equivalent to the standard secular approximation~\cite{Breuer2002,Gardiner2004} but nevertheless yields the desired Lindblad form of the master equation.
Importantly, this second approximation has a different error bound, $\mathcal{E}_L$, than the Wangsness-Bloch-Redfield approach. However, there exists a single quantity, $\mathcal{E}$, which is larger than both $\mathcal{E}_M$ and $\mathcal{E}_L$, which serves as error bound for the effective Lindbladian approximation.

In order to derive \Eq{LindbladForm}, one starts from the Wangsness-Bloch-Redfield approximation which can be written as~\cite{Bloch1946Oct,Wangsness1953Feb,Redfield1965Jan,Breuer2002}
\begin{align}
\dot{\rho}(t) &= \mathcal{D}_R(t) \rho(t) + \mathcal{E}_M,\label{BlochRedfield}
\end{align}
where $\mathcal{E}_M$ is the error introduced by this approximation. The retarded dissipator is given by
\begin{equation}
\mathcal{D}_R(t) = \int_{-\infty}^t dt' \Delta_1(t,t'),
\end{equation}
where the bath memory kernel superoperator, $\Delta_1(t,t')$, is (in the interaction picture) defined by
\begin{equation}
\Delta_1(t,t') \mathcal{O} =- \tr_B[H_I(t),[H_I(t'),\mathcal{O}]].
\end{equation}
Here $\tr_B$ indicates a trace over the bath degrees of freedom.
We note that the Born (weak-coupling) approximation has been used to derive Eq.~\eqref{BlochRedfield}.
The error of the approximation may be bounded as \cite{Albash2012Dec}
\begin{equation}\label{ErrorMarkov}
	\mathcal{E}_M \leq 2g\tilde\Gamma \int_0^\infty dt \, t\abs{B(t)} ,
\end{equation}
with
\begin{equation}
	\tilde\Gamma = g\int_0^\infty dt \, \abs{B(t)}.
\end{equation}
The latter quantity serves as bound for the rate of change of the reduced density matrix
in the maximal eigenvalue norm,
\begin{equation}
	 \vert\vert\dot{\rho}(t)\vert\vert \leq \tilde\Gamma.
\end{equation}
Nathan and Rudner \cite{Nathan2020Apr} also show that the Born approximation alone introduces an error of size $\mathcal{E}_M/2$, and  thus, in a sense, is already equivalent to the full Born-Markov approximation, which accounts for an additional error bounded by $\mathcal{E}_M/2$.
We note that this argument only holds true on short time scales, since small deviations in $\dot{\rho}(t)$ may lead to very different long-time limits.

From the above starting point, one then derives the following bound~\cite{Nathan2020Apr}:
\begin{equation}
\dot{\rho}(t) = \mathcal{L}(t) \rho(t) + \mathcal{E}_M+\mathcal{E}_L.
\end{equation}
This equation is of Lindblad form, see Eq.~\eqref{LindbladForm}, with the single jump operator $L$ in Eq.~\eqref{NLJumpOp} and the rate $\Gamma=1$.
The Lamb shift contribution $H_{LS}$ is discussed in Sec.~\ref{sec3c} below.
The new error term, $\mathcal{E}_L$, is bounded according to
\begin{align}
	\mathcal{E}_L &\leq 2 g \tilde\Gamma\int_{0}^\infty dt dt'  \, t \abs{h(t)}  \abs{h(t')} ,\\
	h(t) &= \frac{1}{2\pi}\int_{-\infty}^\infty d\omega\, \sqrt{B(\omega)} e^{-i\omega t}.\nonumber
\end{align}
Moreover, one finds \cite{Nathan2020Apr}
\begin{align}\label{bound1}
	\mathcal{E}_M, \mathcal{E}_L \leq\mathcal{E},&\quad {\cal E}=\eta\tilde\Gamma,
\end{align}
where we define the dimensionless number
\begin{equation}\label{etaExpr}
	\eta = 2g \int_{-\infty}^\infty dt dt'  \,  \abs{t \, h(t)}  \abs{h(t')}.
\end{equation}
The effective Lindbladian approximation is then justified for $\eta\ll 1$.

We emphasize that the error bound ${\cal E}$ is conservative. Taking, e.g., a thermal bosonic bath, the error bound diverges in the infinite-temperature limit owing to the presence of $\nB(\nu)$ in Eq.~\eqref{eq_thermal_bath_correlation}, even though the Markovian approximation should be valid in this limit. Furthermore, $\mathcal{E}$ tends to be at least an order of magnitude larger than $\mathcal{E}_M$ in the cases considered below. For the numerical results shown in Sec.~\ref{sec3}, we have chosen model parameters in a conservative manner, such that $\mathcal{E}$ is at most comparable to the slowest non-vanishing decay rate of the problem. However, we expect that the effective Lindbladian approximation remains accurate even when less conservative parameters are chosen. Moreover, since $\mathcal{E}\propto g^2$, the error bound can always be made arbitrarily small against the relevant relaxation and decoherence rates by reducing $g$, since those rates already receive contributions $\propto g$.
We discuss the error bound in more detail in Sec.~\ref{sec3b1}.

\section{Results}\label{sec3}

\subsection{Results for generic environments}\label{sec3a}

Making use of the effective Lindbladian approximation, see Eqs.~\eqref{LindbladForm} and \eqref{NLJumpOp}, we obtain an explicit expression for the Liouvillian, $\mathcal{L}$, that holds for an arbitrary bath correlation function $B(\omega)$. Just as the Hamiltonian is a block-diagonal operator, the Liouvillian is a block-diagonal superoperator. We parametrize the reduced density matrix as
\begin{align}
\rho &= \begin{pmatrix} \rho_+ & \rho_c \\ \rho_c^\dagger & \rho_- \end{pmatrix}, \quad\rho_i = \begin{pmatrix} a_i & b_i \\ c_i & d_i \end{pmatrix}, \quad i =\pm, c,\label{BlockDiagBasis}
\end{align}
where the diagonal blocks $\rho_\pm$ refer to the parity $s=\pm 1$ in Eq.~\eqref{subparity}.
Noting that $b_\pm = c_\pm^*$, the time evolution is given by ($i=\pm,c$)
\begin{equation}
\rho_i(t)  = e^{\mathcal{L}_i t} \rho_i(t=0).
\end{equation}
We refer the reader to App.~\ref{AppB} for the explicit form of the superoperators $\mathcal{L}_i$. Their complex-valued eigenvalues, $\{\Lambda^i_j\}$, contain
information about the rate of change in the corresponding density matrix block $i$. Specifically, the respective decay rates are given by
\begin{equation}\label{DecayRatesGen}
\Gamma^i_j = - \Re\Lambda^i_j.
\end{equation}

For the diagonal blocks ($i=s= \pm$), the problem is formally identical to a single spin-boson model, see Sec.~\ref{sec2d}.
There is one zero eigenvalue, $\Lambda_0^s=0$, corresponding to the steady state reached at very long times.
To lowest order in $g$ and using Eq.~\eqref{DecayRatesGen}, we obtain the decay rates describing the approach to the steady state,
\begin{subequations}
\begin{align}
	\Lambda_1^s &= -g\frac{\Delta_s^2}{4E_s^2} \left[ B(E_s) + B( -E_s)\right] +\mathcal{O}(g^3),\label{LambdaDiag1}\\
	\Lambda_{2,\pm}^s &=-\frac{g}{8E_s^2}\left(  \Delta_s^2 \left[B(E_s) + B(-E_s) \right] + 4\epsilon^2 B(0) \right)\notag\\
	&\phantom{=}\quad \pm iE_s + \mathcal{O}(g^2),\label{LambdaDiag2}
\end{align}
\end{subequations}
with $\Delta_s$ in Eq.~\eqref{RotHam} and $E_s$ in Eq.~\eqref{eq_different_energies}.  For a parity readout of the dot-MBS system,
these rates describe how fast the dot charge (or the quantum capacitance) will reach its final value at long times.
The respective density matrix block in this long-time limit is determined by the kernel of $\mathcal{L}_i$.
For the diagonal block with $s=\pm 1$, using the energy eigenbasis \eqref{eq_different_energies}, we obtain
 \begin{equation}
	\rho_s(\infty) = \frac{1}{A_s^++A_s^-} \begin{pmatrix} A_s^- & -i\frac{g}{E_s}A_s^c\\ i\frac{g}{E_s} A_s^c  &
	A_s^+ \end{pmatrix},
\end{equation}
with the quantities
\begin{align}
A_s^\pm &=\frac{\Delta_s^2}{4E_s^2} B(\pm E_s),\quad A_s^c= A_s^- n_s^{+}-A_s^+ n_s^-,\label{DefnAsp}\\ \nonumber
n_{s}^\pm&= \frac{\Delta_s \epsilon}{8 E_s^2}  \sqrt{B(0)}\left(3\sqrt{B(\pm E_s)}-\sqrt{B(\mp E_s)} \right).
\end{align}
In addition, to order $\mathcal{O}(g)$,
the steady-state expectation value for the dot occupation number in block $s=\pm 1$ is given by
\begin{align}
	\av{n_d(\infty)}_s&= \frac{1}{2}\left(1-\av{\sigma_z(\infty)}_s \right)\nonumber\\
	&=\frac{1}{2} \left(1 - \frac{\epsilon}{E_s}\frac{B(E_s) - B(-E_s)}{B(E_s)+B(-E_s)}\right).\label{EquilibriumCharge}
\end{align}
Similarly, the respective saturation value for the quantum capacitance follows as $\frac{d}{d\epsilon}\av{n_d(\infty)}_s$.

Finally, for the coherence block ($i=c$), one finds only non-zero eigenvalues.
Up to order ${\cal O}(g^2)$ terms, with $p_1,p_2=\pm 1$, they are given by
\begin{align}
	\Lambda_{p_1,p_2}^c &= -\frac{g}{2}\left(A_{p_1}^{p_2} + A_{-p_1}^{- p_1 p_2} + 2K^{p_1} \right) + ip_2  f^{p_1},\label{CoherenceRates}
\end{align}
with the quantities $A_s^\pm$ in Eq.~\eqref{DefnAsp} and
\begin{equation}\label{kfdef}
K^\pm = \frac{\epsilon^2 (f^\pm)^2}{2E_+^2 E_-^2} B(0),\quad
f^\pm = \frac{1}{2}(E_+ \pm E_-).
\end{equation}
We now proceed by illustrating these general results for the specific environments in Sec.~\ref{sec2c}.

\subsection{Results for specific environments}\label{sec3b}

One of the main results of this work is stated in Eq.~\eqref{CoherenceRates}, which yields the rates $\Gamma_{p_1p_2}^c=-\Re\Lambda^c_{p_1p_2}$ (with $p_1,p_2=\pm 1$) governing the decay of quantum coherence shared by the two parity subblocks $s=\pm1$, see Eq.~\eqref{subparity}. Along with the (known) relaxation rates for the spin-boson model \cite{Weiss2011Nov}, see Eqs.~\eqref{LambdaDiag1} and \eqref{LambdaDiag2},
these results allow one to obtain explicit estimates for the relaxation and/or decoherence time scales
characterizing the dot-MBQ system coupled to a generic environment with the correlator $B(\omega)$.
In this subsection, we examine these results for the specific environments in Sec.~\ref{sec2c}.

\subsubsection{Thermal bath of bosons}\label{sec3b1}

We begin with a bosonic bath in thermal equilibrium, see Sec.~\ref{sec2c1}. For an Ohmic bath, the correlator $\Bth(\omega)$ is given by Eq.~\eqref{thermalBomega}. We choose an exponential cutoff function, ${\cal C}(\omega,\omega_c)=e^{-|\omega|/\omega_c}$, where $\omega_c$ is the bath cutoff frequency.

First, in order to obtain the dimensionless number $\eta=\eta_{\text{th}}$, we have numerically computed the integrals in Eq.~\eqref{etaExpr}
as a function of $k_BT/\omega_c$.
The error bounds discussed in Sec.~\ref{sec2e} imply the condition $\eta_{\rm th}\ll 1$
for the effective Lindbladian approximation. Within the temperature range
\begin{equation}\label{window}
    0.001\omega_c\alt k_B T\alt 10\omega_c,
\end{equation}
we find $\eta_{\rm th}< 100g$, with a broad minimum at $\eta_{\rm th}\approx 10 g$ around $k_B T\approx 0.1\omega_c$.
For small system-bath couplings, say, $g\alt 0.001$, we conclude that the effective Lindbladian approximation is safely controlled within the temperature window \eqref{window}. The error bound ${\cal E}_{\rm th}=\eta_{\rm th}\tilde\Gamma$, see Eq.~\eqref{bound1}, is then smaller than the predicted decay rates.
The error bound may, however, become larger for either very low or very high temperatures.
The case of very high temperatures has already been discussed in Sec.~\ref{sec2e}.
Moreover, in the zero-temperature limit, one generally expects the Lindblad equation to break down \cite{Breuer2002,Weiss2011Nov}.
However, let us also recall that this error bound is conservative, and the actual error introduced by the effective Lindbladian approximation may in fact be much smaller, see Sec.~\ref{sec2e}. Finally, we note that for the numerical calculation of $\eta_{\rm th}$,
we have used a long-time integration cutoff $t_{\rm max}$ in Eq.~\eqref{etaExpr}, which physically corresponds to the total duration of the measurement.  Sending $t_{\rm max}\to \infty$, one encounters a  weak logarithmic divergence of $\eta_{\text{th}}$, see also Refs.~\cite{Mozgunov2020Feb,Nathan2020Apr}.

For the diagonal blocks ${\cal L}_i$ with $i=s=\pm 1$, we recover from Eqs.~\eqref{LambdaDiag1} and \eqref{LambdaDiag2} the known thermalization rates of the spin-boson model to lowest order in the coupling $g$ \cite{Weiss2011Nov},
\begin{align}
\Gamma_{1,\text{th}}^s &=  \frac{\pi g\Delta_s^2}{4E_s} e^{-E_s/\omega_c} \coth\left(\frac{E_s}{2k_B T}\right), \nonumber\\
\Gamma_{2,\text{th}}^s &= \frac12 \Gamma_{1,\text{th}}^s  + \frac{\pi g}{2} \frac{\epsilon^2}{E_s^2} k_B T.\label{Gammadiag2}
\end{align}
These rates tell us how quickly thermalization occurs, i.e., on which time scales the density matrix of the combined dot-MBS system will approach the thermal state in Eq.~\eqref{Thermalstate}.
Turning to the coherences between the $s=+1$ and $s=-1$ sectors, Eq.~\eqref{CoherenceRates} yields the corresponding four decay rates to order $\mathcal{O}(g)$. With $p_1,p_2=\pm 1$, we find
\begin{align}
\Gamma_{p_1,p_2,{\rm th}}^c &= \frac{\pi g}{8} \Biggl\{\frac{\Delta_{p_1}^2}{2E_{p_1}} e^{-\frac{E_{p_1}}{\omega_c}}\left[\coth\left(\frac{E_{p_1}}{2k_B T}\right) + p_2 \right]\notag\\
&+ \frac{\Delta_{-p_1}^2}{2E_{-p_1}} e^{-\frac{E_{-p_1}}{\omega_c}}\left[\coth\left(\frac{ E_{-p_1}}{2k_B T}\right) - p_1 p_2 \right]\notag\\
&+ \frac{\epsilon^2 (E_+ + p_1 E_-)^2}{E_+^2 E_-^2} k_B T\Biggr\}.\label{Gammacoh1}
\end{align}
Three of these rates approach a finite value as $T\to 0$ and therefore describe parity thermalization of the coupled dot-MBS system.
However, the smallest rate, $\Gamma_{1,{\rm th}}^c\equiv \Gamma_{+,-,{\rm th}}^c$, vanishes in the $T\to 0$ limit and corresponds to a dephasing rate for
inter-parity quantum coherence.
  From Eq.~\eqref{Gammacoh1},  we find the low-temperature behavior
\begin{equation}\label{ThermalDecRateSmallT}
	\Gamma_{1,\text{th}}^c (T\to 0) \simeq \frac{\pi g}{8}  \frac{\epsilon^2 (E_+ - E_-)^2}{E_+^2 E_-^2 } k_B T .
\end{equation}

\begin{figure}
\includegraphics[width=\linewidth]{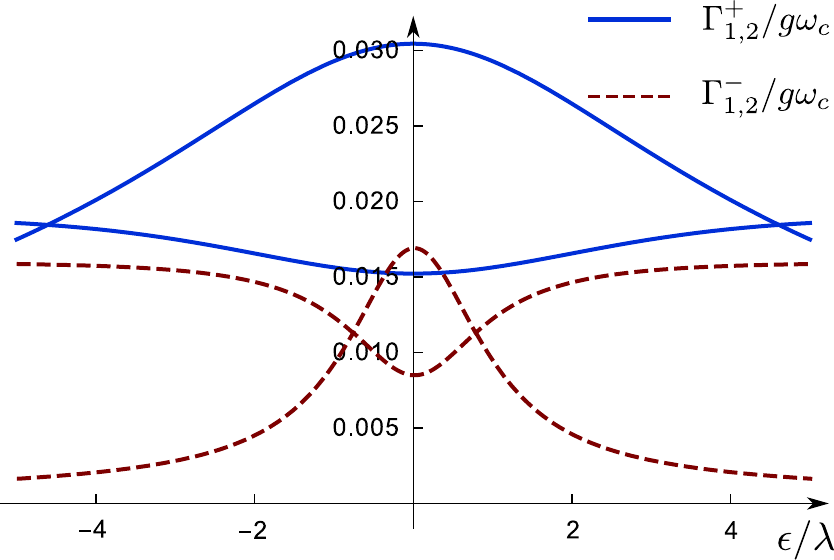}
\caption{Thermalization rates $\Gamma_{1/2,{\rm th}}^{s}$ (in units of $g\omega_c$), see Eq.~\eqref{Gammadiag2}, vs $\epsilon/\lambda$ for a thermal boson bath.
These rates describe thermalization of the diagonal density matrix blocks with parity $s=\pm 1$, where blue solid (red dashed) curves are for $s=+1$ ($s=-1$).
We use the parameters $k_BT=\lambda = 0.01\omega_c$, $a=1$, and $\phi=\pi/3$.
}\label{fig5}
\end{figure}

\begin{figure}
\includegraphics[width=\linewidth]{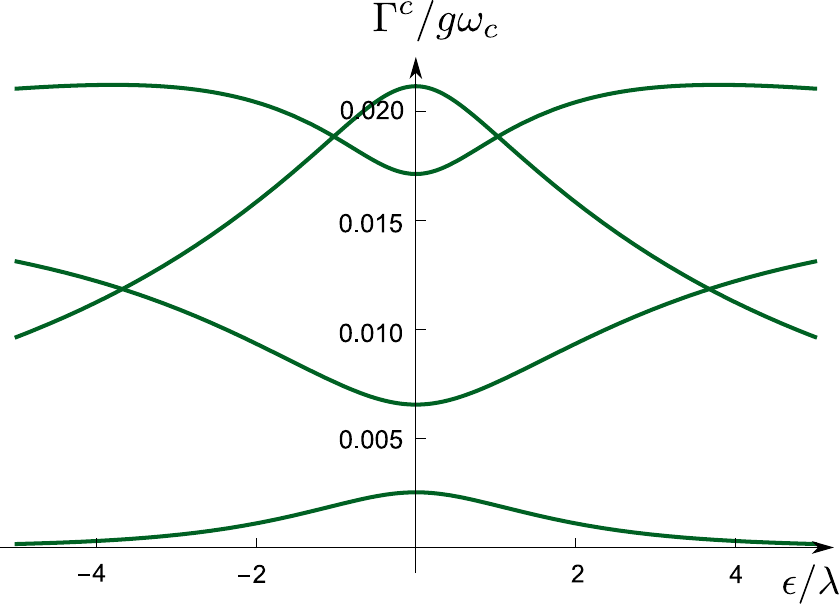}
\caption{Rates describing the decay of quantum coherence between different parity sectors, see Eq.~\eqref{Gammacoh1}, for a thermal bosonic bath. We show the four
rates $\Gamma^c_{p_1p_2,{\rm th}}$ (in units of $g\omega_c$), see Eq.~\eqref{Gammacoh1}, vs $\epsilon/\lambda$, for the parameters in Fig.~\ref{fig5}.
}\label{fig6}
\end{figure}

\begin{figure}
\centering
\includegraphics[width=\linewidth]{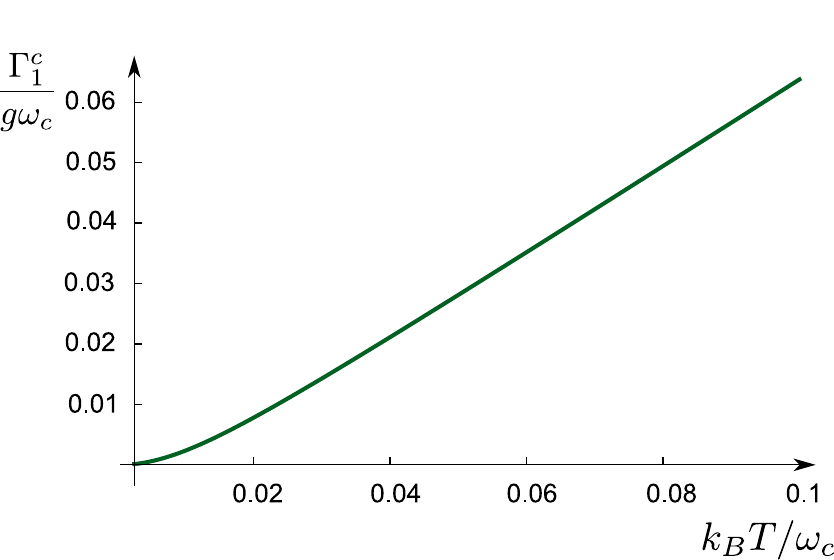}
\caption{Inter-parity dephasing rate, $\Gamma_{1,{\rm th}}^c=\Gamma_{+-,{\rm th}}^c$ (in units of $g\omega_c$), vs $k_BT/\omega_c$ for a thermal bosonic bath,
see Eq.~\eqref{Gammacoh1}. We use the
parameters in Figs.~\ref{fig5} and \ref{fig6} with $\epsilon=\lambda/2$.  The low-temperature behavior is given by Eq.~\eqref{ThermalDecRateSmallT}.   }\label{fig7}
\end{figure}

Figures \ref{fig5}, \ref{fig6}, and \ref{fig7} illustrate the above results.
The decay rates in the diagonal sector, see Eq.~\eqref{Gammadiag2}, are shown in Fig.~\ref{fig5}, while
the decay of quantum coherence between the two parity sectors is shown in Fig.~\ref{fig6}, see Eq.~\eqref{Gammacoh1}, and in Fig.~\ref{fig7}.
In Figs.~\ref{fig5} and \ref{fig6}, we show the respective rates at fixed temperature
as a function of the ratio $\epsilon/\lambda$ between the dot level energy $\epsilon$ and the overall tunneling strength $\lambda$.
We observe that some of the inter-parity decay rates are of the same order
of magnitude as the thermalization rates in the parity-diagonal sectors.  These inter-parity rates also do not vanish in the $T\to 0$ limit and correspond to thermalization rates of the system.
In the long-time limit, the smallest of the rates shown in Fig.~\ref{fig6} dominates the approach to the steady state.
The dephasing rate in the off-diagonal parity sector, $\Gamma_{1,{\rm th}}^c$, is shown in Fig.~\ref{fig7} and vanishes according to Eq.~\eqref{ThermalDecRateSmallT} as
$T\to 0$.  Our results show that quantum coherence between different parity sectors can persist for long time scales at  low
temperatures.

We also note that in the long-time limit, the expectation value of the dot occupation number approaches the thermal equilibrium value.
Indeed, Eq.~\eqref{EquilibriumCharge} yields
\begin{equation}\label{avnd1}
	\av{n_d(\infty)}_{s,{\rm th}} = \frac{1}{2}\left[1-\frac{\epsilon}{E_s} \tanh\left(\frac{E_s}{2 k_B T}\right)\right].
\end{equation}
This result holds for arbitrarily small (but finite) $g$.

As concrete example for a thermal bosonic bath, we now consider the electromagnetic environment corresponding to the circuit in
Fig.~\ref{fig3}, where the bath correlator has been specified  in Eq.~\eqref{BLC}.
In effect, the respective decay rates can then be inferred from the above results by replacing
\begin{equation}
    g e^{-\frac{E_s}{\omega_c}} \to g_{LC}\, \mathcal{C}\left(E_s/\omega_{LC}\right),
\end{equation}
with the Lorentzian cutoff function $\mathcal{C}(\tilde \omega)$ in Eq.~\eqref{w12}. The coupling $g_{LC}$ and the $LC$ resonance frequency $\omega_{LC}$ have been
specified in Eq.~\eqref{glcwc}.
For instance, the first of the two thermalization rates in Eq.~\eqref{Gammadiag2}, for the diagonal sector with parity $s=\pm 1$, is given by
\begin{equation}
	\Gamma_{1,LC}^s = \frac{\pi g_{LC} \Delta_s^2}{4E_s} \mathcal{C}\left(\frac{E_s}{\omega_{LC}}\right) \coth\left(\frac{E_s}{2k_BT}\right),
\end{equation}
with $\Delta_s$ in Eq.~\eqref{RotHam}. Similarly,
we find from Eq.~\eqref{ThermalDecRateSmallT} the low-temperature behavior of the dephasing rate for inter-parity quantum coherence,
\begin{equation}
	\Gamma_{1,LC}^c\simeq \frac{g_{LC}}{4}
	\frac{\omega_{LC}}{\omega_1+\omega_2}  \frac{\epsilon^2 (E_+ - E_-)^2}{E_+^2 E_-^2 } k_B T ,
\end{equation}
with $\omega_{1,2}=1/(R_{1,2}C)$.

\subsubsection{QPC detector}\label{sec3b2}

We next turn to the nonequilibrium environment corresponding to the QPC measurement setup shown in Fig.~\ref{fig4}, see Sec.~\ref{sec2c2}.
For this QPC charge readout of the parity of the dot-MBQ state, the bath correlator is given by \Eq{eq_qpc_bath_correlation_fourier_final}.
For simplicity, we focus on the effect of the potential gradient $\Delta\mu=\mu_L-\mu_R>0$ across the QPC, and neglect the purely thermal contribution to $B_{\rm QPC}(\omega)$,
which on its own has already been studied in Sec.~\ref{sec3b1}.
For $\Delta\mu\ll \omega_c$, we thus take the bath correlator responsible for the QPC charge readout as
\begin{align}
	B_{{\rm QPC}} (\omega) &\simeq \frac{\pi }{2}\sum_{p=\pm}(\omega + p \Delta \mu) e^{-\abs{\omega+p\Delta\mu}/\omega_c}\notag\\
	&\times \left[\coth\left(\frac{\omega+p\Delta \mu}{2k_BT}\right) +1\right].\label{BQPC}
\end{align}
In this case, our numerical analysis of Eq.~\eqref{etaExpr} shows that with increasing potential bias $\Delta\mu$, the
parameter $\eta$ becomes smaller.  In a sense, the bias $\Delta\mu$ acts like an effective temperature and by increasing its value, the memory time
of the bath becomes shortened \cite{Altland2009Jan}.  For example, using $\Delta\mu=0.1\omega_c$ and $g=0.001$, we find that in contrast to Eq.~\eqref{window},
the effective Lindbladian approximation stays accurate for all temperatures $k_B T\alt 10\omega_c$, down to zero temperature.

\begin{figure}
\centering
\includegraphics[width=\linewidth]{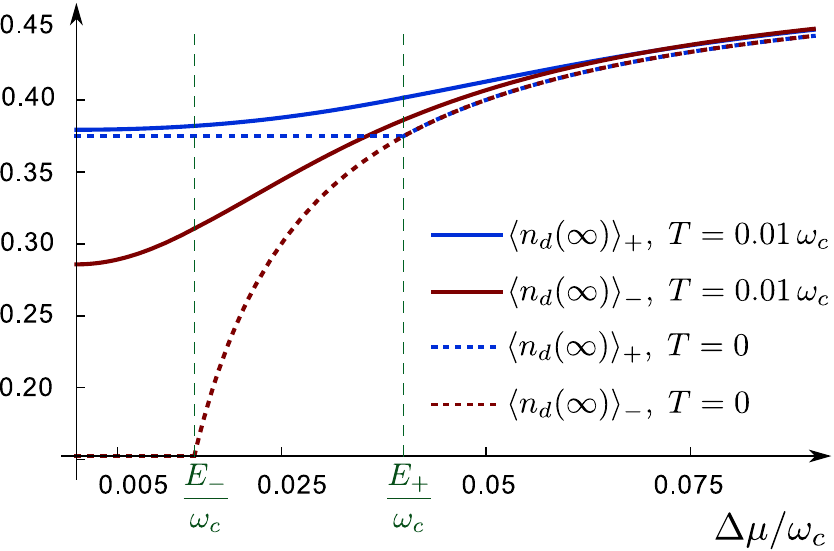}
\caption{Steady-state dot occupation number, $\av{n_d(\infty)}_s$, vs potential bias $\Delta\mu$
for the QPC parity readout with $s=\pm 1$, see Sec.~\ref{sec3b2}, obtained from Eq.~\eqref{EquilibriumCharge}
with Eq.~\eqref{BQPC}, $\epsilon =0.01\omega_c, \lambda=\epsilon, a=1$, and $\phi=\pi/3$.
The system energies $E_{\pm}$ in Eq.~\eqref{eq_different_energies} are shown as green vertical dashed lines.
Blue (red) curves are for parity $s=+1$ ($s=-1$).
Solid curves are for $k_BT =0.01\omega_c$.
The corresponding analytical $T=0$ results, Eq.~\eqref{ndT0},
are shown as dashed curves.
}\label{fig8}
\end{figure}

\begin{figure}
\centering
\includegraphics[width=\linewidth]{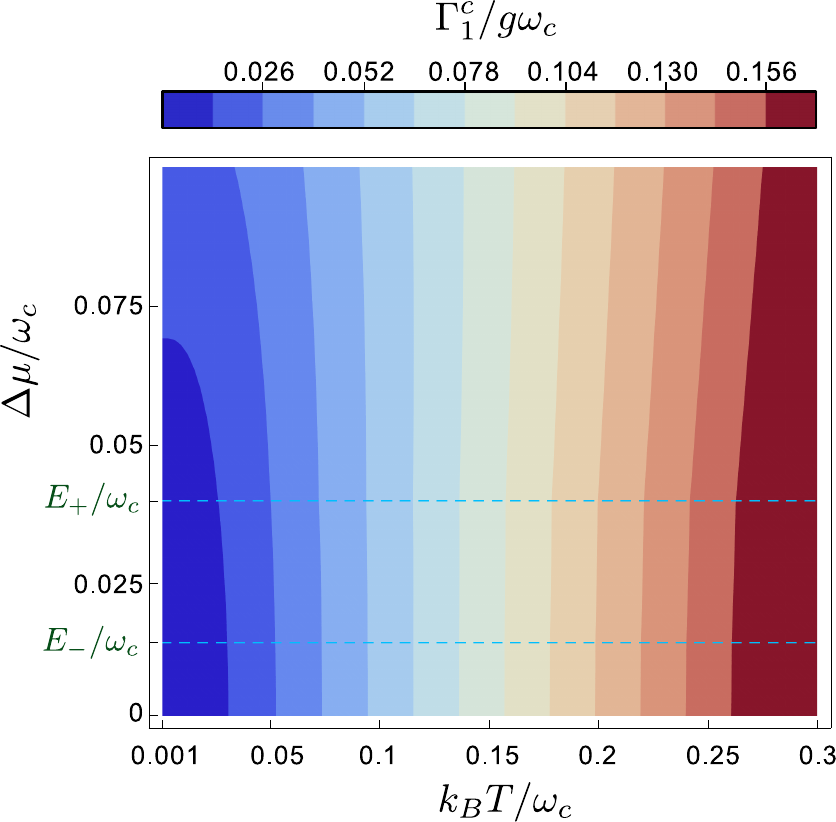}
\caption{Dephasing rate for parity off-diagonal quantum coherence, $\Gamma_1^c$ (in units of $g\omega_c$), in the $T$-$\Delta\mu$ plane, for the case of a
QPC readout environment with $\epsilon=\lambda= 0.01\omega_c, a=1,$ and $\phi = \pi/3$.
The dashed horizontal lines correspond to $\Delta\mu=E_{s=\pm}$, see Eqs.~\eqref{eq_different_energies} and \eqref{ndT0}.  }
\label{fig9}
\end{figure}

Let us now turn to the zero-temperature limit in order to study how decoherence in our system will depend on $\Delta\mu$.
For $T= 0$ and $0<E_s,\Delta\mu\ll \omega_c$, Eq.~\eqref{BQPC} simplifies to
\begin{equation}
	B_{{\rm QPC},T=0}(\omega)\approx \pi  \sum_{p=\pm} \left|\omega + p \Delta \mu\right| \Theta(\omega +p \Delta\mu),
\end{equation}
where $\Theta(x)$ is the Heaviside step function. Moreover, from Eq.~\eqref{EquilibriumCharge}, we obtain the
average steady-state  dot occupation number as
\begin{equation}\label{ndT0}
	\av{n_d(\infty)}_s =\begin{cases}  \frac12(1- \epsilon/E_s), & \Delta \mu < E_s, \\ \frac12(1- \epsilon/\Delta\mu), & \Delta \mu \geq E_s.\end{cases}
\end{equation}
In order to read out the parity $s=\pm 1$, we evidently cannot have $\Delta \mu\geq E_s$ for both values of $s$. On the other hand, if $\Delta \mu < E_s$ for both $s$, the dependence on $\Delta\mu$ drops out completely, resulting in the optimal case of maximum visibility.
We illustrate the average steady-state dot occupation number in Fig.~\ref{fig9}, where we observe that
while the above $T=0$ argument basically carries over to the finite temperature case, the sharp changes at $\Delta\mu=E_s$ in Eq.~\eqref{ndT0} are smeared
out by thermal fluctuations.

The smallest non-vanishing decay rate at $T=0$ in the diagonal block with parity $s=\pm 1$ is then given by
\begin{equation}\label{QPCrate1}
	\Gamma_1^s =  \frac{\pi g}{4} \frac{\Delta_s^2}{E_s^2} \times \begin{cases} E_s, &\Delta\mu< E_s, \\ \Delta \mu,& \Delta \mu \geq E_s. \end{cases}
\end{equation}
For the optimal visibility case with $\Delta\mu <E_s$ for both values of $s$, this result formally coincides with the smallest
thermal rate at zero temperature, see Eq.~\eqref{Gammadiag2}.
Importantly, the decay rate is then insensitive to the value of the potential bias $\Delta \mu$. For the decay of the off-diagonal coherences,
we find that the $T=0$ dephasing rate, $\Gamma_1^c(\Delta\mu)$, depends linearly on the potential bias for $\Delta\mu<E_\pm$,
\begin{equation}
	\Gamma_1^c(T=0,\Delta\mu) \simeq \frac{\pi g}{16}   \frac{\epsilon^2(E_+ - E_- )^2}{E_+^2E_-^2}  \Delta\mu.
\end{equation}
By comparing this result to the thermal rate in Eq.~\eqref{ThermalDecRateSmallT}, we observe that the potential bias plays the role of
an effective temperature, as expected on general grounds \cite{Altland2009Jan}.
In the opposite limit, $k_B T \gg \Delta\mu$, the dephasing rate is basically described by the results in Sec.~\ref{sec3b1}.

Figure \ref{fig9} illustrates the dephasing rate $\Gamma_1^c$ for the case of a QPC detector, as a function of both temperature and bias voltage.  These results were obtained from Eq.~\eqref{CoherenceRates}.  We first observe that at low temperatures, the dephasing rate increases with increasing potential bias.  This behavior is expected because the potential bias acts as effective temperature.  On the other hand, for $k_B T\agt \Delta\mu$, the potential bias has little effect on the rate which now is dominated by thermal fluctuations.
Next, we note that in the potential bias window where different parity states can be distinguished with good visibility, the time scale $\tau_M=1/\Gamma_1^c$ for the off-diagonal coherence to decay, and hence the time it takes to make a projective measurement of the parity $s=\pm 1$, is limited to a time of the order $\tau_M\approx 10^6/\omega_c$. On the other hand, the readout time is determined by $\tau_R= \min_{s} 1/\Gamma_{1}^s$, i.e., in terms of the decay rates in the diagonal sector.
Now $\tau_R$ is typically shorter than $\tau_M$, which implies that if the system parameters are chosen such that the final state allows one to distinguish the two values of $s$, the time $\tau_M$ will effectively determine the readout time of the measurement.
Finally, we note that from Fig.~\ref{fig8}, one observes that for good visibility, one needs $\Delta\mu \lesssim E_+$.  This observation
suggests that a readout procedure with an initially larger value of $\Delta\mu$ may be advantageous since in this manner one can speed up the off-diagonal decay.
Subsequently using a smaller potential bias $\Delta \mu$, one can then maximize visibility.

\subsection{Lamb shift}\label{sec3c}

The Lamb shift can be thought of as a renormalization of the dot-MBQ energies by the bath modes. This renormalization does not contribute to decay rates but contributes to the effective Hamiltonian appearing in the Liouvillian. So far we have not discussed the corresponding term, $H_{LS}$, which appears in the coherent time evolution part of Eq.~\eqref{LindbladForm}. The Lamb shift could potentially be important for the readout, for instance, by reducing the visibility in the readout via $s$-dependent shifts of the average dot occupation $\av{n_d}_s$.

In this subsection, we show that for $E_\pm \ll \omega_c$,
$H_{LS}$ only causes an $s$-independent constant energy shift.  As a consequence, the Lamb shift
is not expected to affect the readout visibility for our dot-MBQ setups.

In the eigenbasis of $H_0$, defined by $H_0 \ket{p,s} = \left( \frac{\epsilon}{2}+\frac{p}{2}E_s\right)\ket{p,s}$ for $p=\pm 1$, see Eq.~\eqref{eq_different_energies}, the Lamb shift in the effective Lindbladian approximation takes the form \cite{Nathan2020Apr}
\begin{align}\label{HLSDefn}
& H_{LS} =  \frac{g}{16} \sum_{p,q,r,s=\pm 1} Z_{pq,s} Z_{qr,s} \\ \nonumber
&\quad \times\left[Q\left(\frac{p-q}{2}E_s\right) + Q\left(\frac{q-r}{2}E_s\right)\right]\outprod{p,s}{r,s},\\
&\quad Q(\omega) = \frac{ \mathcal{P} }{\pi}\int_{-\infty}^\infty d\nu\frac{B(\nu)}{\omega - \nu}, \nonumber
\end{align}
where ${\cal P}$ denotes the principal part of the integral and $B(\nu)$ is the bath correlator for the respective environment. We employ the quantities
 \begin{equation}
	Z_{pq,s} \equiv \inprod{p,s}{\sigma_z}{q,s} = \begin{cases}-p \epsilon/E_s, & p=q,\\  \Delta_s/E_s, & p=-q,\end{cases}
\end{equation}
with $\Delta_s$ in Eq.~\eqref{RotHam},
such that Eq.~\eqref{HLSDefn} can be written as
\begin{equation}\label{LambShiftGeneral}
	H_{LS} = \frac{g}{8}\sum_{s}\left(\frac{ \epsilon^2Q(0)}{E_s^2} +\frac{\Delta_s^2[Q(E_s)+Q(-E_s)]}{2E_s^2} \right)\Pi_s,
\end{equation}
where $\Pi_s$ is the projector onto the diagonal parity block with $s=\pm 1$.  The Lamb shift therefore shifts the energies in each block.

We next discuss the form of $H_{LS}$ for the different environments introduced above.
Using Eq.~\eqref{LambShiftGeneral} and symmetry relations obeyed by $B(\omega)$ corresponding to Eq.~\eqref{autocor},
we find that the Lamb shift $H_{LS}$ is independent of temperature. Crucially, for $E_s\ll \omega_c$, we will show that the
energy shift is $s$-independent for all these cases, and therefore it indeed is irrelevant with respect to the parity readout.
 The Lamb shift is also negligible with regard to the average dot occupation $\av{n_d(\infty)}_s$, since Eq.~\eqref{avnd1} is
already determined by contributions of order ${\cal O}(g^0)$.

\subsubsection{Thermal boson bath}

We first evaluate Eq.~\eqref{LambShiftGeneral} for thermal bosons. For an Ohmic bath, using the bath correlator $\Bth(\omega)$ in Eq.~\eqref{thermalBomega}
with an exponential cutoff function,  Eq.~\eqref{LambShiftGeneral} yields the result
\begin{equation}\label{hls1}
H_{LS} = \frac{g\omega_c}{8}  \sum_{p,s=\pm}\left[1 +  \frac{ \Delta_s^2}{2E_s^2}\, \xi\left( \frac{E_s}{\omega_c} \right) \right] \outprod{p,s}{p,s},
\end{equation}
where $\xi(x) = xe^{x}\, {\rm Ei}(-x)-x e^{-x} \, {\rm Ei}(x)$
with the exponential integral, ${\rm Ei}(x)= -\int_x^\infty dt e^{-t}/t$.
Using $\xi(x) \to 0$ for $x \to 0$, we find that for $E_\pm\ll \omega_c$, Eq.~\eqref{hls1} reduces to the constant energy shift $g \omega_c/8$ which
does not affect the parity readout.

Next we turn to  the electromagnetic environment in Fig.~\ref{fig3}, with the bath correlator $B_{LC}(\omega)$ in Eq.~\eqref{BLC}, where the Lamb shift takes a more complicated form. Using the cutoff function ${\cal C}(\tilde\omega)$ in Eq.~\eqref{BLC} with $\omega_c=\omega_{LC}$,  we find
\begin{align}\nonumber
	Q(0) &= \frac{e^2 }{4C} \mathcal{P} \int_{-\infty}^\infty d\tilde{\omega}\, {\cal C}(\tilde{\omega})\\
	Q(E_s)+Q(-E_s)&= \frac{e^2}{4C} \mathcal{P} \int_{-\infty}^\infty d\tilde{\omega}\, {\cal C}(\tilde{\omega}) \\
	&\times\left(\frac{\tilde{\omega}}{\tilde{\omega}+\frac{E_s}{\omega_{LC}}} +
	\frac{\tilde{\omega}}{\tilde{\omega}- \frac{E_s}{\omega_{LC}}}  \right).\notag\nonumber
\end{align}
Using these expressions, we observe that the $s$-dependence drops out again in $H_{LS}$ in the parameter regime $E_\pm\ll \omega_c=\omega_{LC}$.

\subsubsection{Lamb shift for QPC}

For the QPC case, we find the Lamb shift
\begin{align}\nonumber
H_{LS} &= \frac{g\omega_c}{8} \sum_{p,s}\left(1 + \frac{ \epsilon^2}{E_s^2} \xi\left( \frac{\Delta\mu}{\omega_c} \right)+
\frac{ \Delta_s^2}{E_s^2} \xi\left( \frac{E_s}{\omega_c} \right) \right)\notag\\
&\phantom{=\frac{1}{8} g \omega_c\sum_{p,s}}\times\outprod{p,s}{p,s},
\end{align}
where $\xi(x)$ has been defined after Eq.~\eqref{hls1}. As in the thermal case, in the limit $E_\pm,\Delta\mu\ll \omega_c$, the Lamb shift has no consequences for the
parity readout.

\section{Outlook}\label{sec4}

The model we have introduced provides a flexible framework, which may be adapted to study other experimental setups and dephasing mechanisms related to the parity-charge conversion process, see also Refs.~\cite{Aseev2018Oct,Szechenyi2019Sep}. Below we sketch possible extensions of our work that we find particularly interesting. However, a more detailed study of these points goes beyond the scope of this paper.

\subsection{Dispersive readout}

One could use our framework to model the effect of dispersive readouts of Majorana qubits \cite{Plugge2017Jan,Grimsmo2019Jun}. To that end, we consider the electromagnetic environment shown in Fig.~\ref{fig3}. To include the effects of the dispersive readout, however, one should explicitly include the driving fields into the model for the environment.  This step  will modify $B(t)$ significantly, leading to dephasing already at zero temperature. From this point on, our approach should then be applicable again. In particular, by calculating $\av{n_d(\infty)}_s$, one can obtain the impedance shift of the system, from which the resulting amplitude and phase shifts of the reflected signal corresponding to the values $s=\pm1$ can be deduced.

\subsection{Other dephasing mechanisms}

Above, we have studied dephasing caused by the measurement circuit during the MBQ readout. In this subsection, we describe how intrinsic sources of dephasing can be included in the formalism.  In particular, we discuss how the time evolution of the density matrix will be changed due to residual Majorana overlap integrals and/or because of quasiparticle poisoning effects.

When allowing for quasiparticles to relax to or be excited from the zero-energy MBS sector, we need, because of total parity conservation, an additional quantum number describing whether the quasiparticle sector has even or odd occupancy. The total parity of the MBSs and the quantum dot is given by
\begin{equation}\label{pdef}
 p= -i\gamma_1\gamma_2\gamma_3\gamma_4\eta_1\eta_2,
\end{equation}
such that $p=\pm1$ is the quantum number that keeps track of whether the quasiparticle number parity has changed. We can then define MBQ Pauli operators $\mathbf{s}=(s_x,s_y,s_z)$ as
\begin{equation}\label{qubitpaulis}
\begin{aligned}
  s_x&=\gamma_1\gamma_3\eta_1\eta_2=i\gamma_2\gamma_4p,\\
   s_y&=\gamma_1\gamma_4\eta_1\eta_2=-i\gamma_2\gamma_3p,\\
   s_z&=\gamma_1\gamma_2\eta_1\eta_2=i\gamma_3\gamma_4p.
\end{aligned}
\end{equation}
In a similar way, we can write the original Pauli operators $\tilde\sigma_\alpha$, see Eq.~\eqref{pauli0}, as
\begin{equation}
\begin{aligned}\label{sigmapaulis}
  \tilde\sigma_x&=-i\gamma_1\eta_2=\gamma_2\gamma_3\gamma_4\eta_1 p,\\
  \tilde\sigma_y&=i\gamma_1\eta_1=\gamma_1\gamma_3\gamma_4\eta_1 p,\\
  \tilde \sigma_z&=-i\eta_1\eta_2=p\gamma_1\gamma_2\gamma_3\gamma_4.
\end{aligned}
\end{equation}
The two sets of Pauli operators commute, $[s_\alpha,\tilde\sigma_{\alpha'}]=0$ for all $\alpha,\alpha'$.

\subsubsection{Majorana overlaps}

Dephasing of a Majorana qubit due to finite MBS overlaps has been studied before by Knapp \emph{et al.}~\cite{Knapp2018Mar}. The Majorana overlaps introduce a Hamiltonian term of the form
\begin{equation}\label{Majoranaoverlap}
  H_{\mathrm{overlap}}=  \sum_{i<j} t_{ij} i\gamma_i\gamma_j=  \mathbf{s} \cdot [ p \mathbf{d}_1+\tilde\sigma_z \mathbf{d}_2],
\end{equation}
where the real-valued vectors $\mathbf{d}_1=(t_{24},-t_{23},t_{34})$ and $\mathbf{d}_2=(t_{13},t_{14},t_{12})$ contain the overlap matrix elements $t_{ij}$.
We observe that the MBS overlaps basically cause the Bloch vector of the MBQ to precess around an axis defined by the vectors $\mathbf{d}_1$ and $\mathbf{d}_2$.
It is straightforward to include Eq.~\eqref{Majoranaoverlap} in the coherent part of the Liouvillian, see Eq.~\eqref{LindbladForm}.
For a detailed discussion of the resulting physics, see Ref.~\cite{Knapp2018Mar}.

\subsubsection{Quasiparticle poisoning}

We now consider quasiparticle poisoning caused by excitations out of the MBS ground state sector and/or by the relaxation of thermally generated quasiparticles into the MBS sector. We will assume that the time scales for these two processes are slow, in particular much slower than relaxation within the quasiparticle continuum. Moreover, the time scale for the spatial equilibration of quasiparticles is also assumed to be much shorter than the typical time between subsequent poisoning events.
These two assumptions imply that the quasiparticle distribution function is identical for all MBS positions.
The Hamiltonian that describes the coupling between quasiparticles and MBSs is then given by \cite{Munk2019Apr}
\begin{gather}
    H_{\qp} = H_F+H_B+H_\pois, \quad H_F = \sum_k E_{k}^\nodag\alpha_{k}^\dag\alpha_{k}^\nodag,\notag\\
    H_B = \sum_{q}\omega_q b_q^\dag b^\nodag_q,\quad H_\pois =  \sum_{i=1}^{4}\gamma_i\sum_{qk} \Gamma_{iqk}\varphi_q,\notag\\
    \Gamma_{iqk}=v_{iqk}^\nodag\alpha_{k}^\nodag-v_{iqk}^*\alpha_{k}^\dag,\quad \varphi_q=b_q^\nodag+b_q^\dag,\label{qp}
\end{gather}
where $\alpha_k$ are fermionic annihiliation operators for above-gap Bogoliubov quasiparticles with energy $E_{k}$.  Moreover, $b_q$ are annihilation operators for bosonic modes (phonons and/or electromagnetic modes) which mediate the coupling between the two fermionic subsystems, $\omega_q$ are boson energies, and $v_{iqk}$ are the coupling matrix elements.
A key point is now that the quasiparticles have different distribution functions depending on the total quasiparticle number being even or odd. Of course, this statement only holds true for a finite system where parity is conserved, but for closed MBQs, this is indeed the case. The difference between the even and odd quasiparticle number sectors is only significant for temperatures $T\lesssim T^*$, where $T^*$ is the characteristic temperature at which the probability of having a single quasiparticle on the island approaches unity. This cross-over temperature is inversely proportional to the volume $V_S$ of the superconductor and given by \cite{Lafarge1993Feb,Tuominen1993May,Higginbotham2015Sep,Munk2019Apr}
\begin{equation}
    T^*\approx \frac{\Delta}{k_BN_\mathrm{eff}},\quad N_\mathrm{eff}=d_S V_S\sqrt{2\pi k_B T\Delta},
\end{equation}
where $d_S$ is the density of states and $\Delta$ the pairing gap.

To take total parity conservation into account, we project the Hamiltonian \eqref{qp} onto the sector with (say) total even occupancy, $H_{\qp}\rightarrow P_e H_\qp P_e$, where $P_e$ is the projection operator to total even parity. We also define separate projection operators for quasiparticles and MBSs onto the respective even and odd parity sectors, $P_{e/o}^{\qp,M}$. With $P_e=P_e^MP^\qp_e+P_o^MP^\qp_o$, the projected poisoning Hamiltonian  becomes
\begin{align}\label{HpoisPeo}
  P_e H_\pois P_e  &= \sum_{i=1}^{4} P_o^M\gamma_i P_e^M \sum_{qk} P_o^\qp\Gamma_{iqk}\varphi_qP_e^\qp\notag\\
  &+\sum_{i=1}^{4} P_e^M\gamma_i P_o^M \sum_{qk} P_e^\qp\Gamma_{iqk}\varphi_q P_o^\qp.
\end{align}
We can now identify two contributions in Eq.~\eqref{HpoisPeo}. The first term couples the MBQ via the operator $\gamma_{i,e\rightarrow o}=P_o^M\gamma_i P_e^M$ to a reservoir with an even number of quasiparticles, while the second term couples it via $\gamma_{i,o\rightarrow e}=\gamma_{i,e\rightarrow o}^\dagger$ to  a reservoir with odd quasiparticle number.  Equation~\eqref{HpoisPeo} allows us to directly apply the effective Lindbladian approximation introduced in Sec.~\ref{sec2e}. To that end, we define a jump operator for each of the two terms in  Eq.~\eqref{HpoisPeo},
\begin{subequations}
\begin{align} \label{LL1}
  L_{e\rightarrow o} =& \sum_i \sum_{mn} \matelem{m}{\gamma_{i,e\rightarrow o}}{n} \sqrt{g_{ii}^e(E_n-E_m)}\ket{m}\bra{n},\\ \label{LL2}
  L_{o\rightarrow e} =& \sum_i \sum_{mn} \matelem{m}{\gamma_{i,o\rightarrow e}}{n} \sqrt{g_{ii}^o(E_n-E_m)}\ket{m}\bra{n},
\end{align}
\end{subequations}
where the two bath functions are given by
\begin{equation}\label{geodef}
  g_{ij}^{e/o}(t)=-\sum_{qk}\average{\Gamma_{iqk}(t)\Gamma_{jqk}(0)}_{e/o}\average{\varphi_q(t)\varphi_q(0)}.
\end{equation}
The fermionic expectation value is here taken over quasiparticle distributions in the respective sector with even or odd total occupation number.
The functions \eqref{geodef} have also been discussed in Refs.~\cite{Higginbotham2015Sep,Lafarge1993Feb,Tuominen1993May,Munk2019Apr,Rainis2012May,Schmidt2012Aug}.
Note that in Eqs.~\eqref{LL1} and \eqref{LL2} we have neglected coherent transport of quasiparticles between the ends of the topological superconductors. If coherent quasiparticle transfer between the wire ends is important, it can be included by creating jump operators from the square roots of the matrices $g_{ij}^{e/o}(\omega)$ \cite{Nathan2020Apr}.

As  final step, we now use the fact that because of the coupling to incoherent quasiparticle reservoirs,  the total even and odd ($p=\pm 1$) sectors of the MBQ have no quantum-coherent coupling.
We can therefore write the dynamical equations for the MBQ reduced density matrices with even or odd parity, $\rho_{e/o}$, as
\begin{subequations}
\begin{align}
  \dot{\rho}_e = & (\dot{\rho}_e)^{(0)} -\frac12 \left\{L_{e\rightarrow o}^\dag L_{e\rightarrow o}, \rho_e\right\} + L_{o\rightarrow e}\rho_o L^\dag_{o\rightarrow e},   \\
  \dot{\rho}_o =& (\dot{\rho}_o)^{(0)}-\frac12 \left\{L_{o\rightarrow e}^\dag L_{o\rightarrow e}, \rho_o\right\} + L_{e\rightarrow o}\rho_e L^\dag_{e\rightarrow o},
\end{align}
\end{subequations}
where $(\dot{\rho}_{e/o})^{(0)}$ is the time derivative in the absence of quasiparticle poisoning.
Finally, we note that the coupling of the MBS sector to the quasiparticle reservoirs will also give rise to Hamiltonian corrections of the same form as the residual overlaps in Eq.~\eqref{Majoranaoverlap}.

\section{Conclusions}
\label{sec5}

We have developed a flexible theory for calculating the thermalization and dephasing rates for arbitrary quantum states of a Majorana box qubit tunnel-coupled to a quantum dot for
parity readout. Our analysis shows
that this parity-to-charge conversion process sensitively depends on the choice of the readout device connected to the dot charge.  The latter can be thought of as a generic Markovian bosonic environment (heat bath), either in thermal equilibrium or operated under nonequilbrium conditions.
Particular care has been taken to properly account for the decay of coherences among blocks with different fermion number parity $s=\pm 1$, where $s$ refers to the parity of the quantum dot together with the two tunnel-coupled Majorana states.

By employing a recently developed effective Lindbladian approximation,
the resulting quantum master equation is by construction of Lindblad form, meaning that complete positivity of the density matrix is guaranteed during the entire time evolution. We have provided explicit results for decay rates when the environment consists of a generic thermal boson heat bath.  An important special case is defined by the electromagnetic fluctuations in a macroscopic electric circuit connected to the Majorana qubit.
In addition, we have examined the nonequilibrium environment corresponding to a Majorana parity readout via conductance measurements of a quantum point contact that is
capacitively coupled to the dot. For all these examples, we have derived analytical expressions for decay rates, which in turn can be related to experimentally measurable quantities. By taking into account quasiparticle poisoning and Majorana overlap effects as sketched in Sec.~\ref{sec4}, it stands to reason that this theoretical approach can  allow for a realistic and powerful description of quantum decoherence in Majorana box qubits.

\emph{Note added:} After completion of this manuscript, we were informed of a closely related independent manuscript by Steiner and von Oppen \cite{Steiner2020Apr}. Their conclusions are consistent with our findings.  Despite of the overlap between both works, they are largely complementary. While we employ the improved jump operators introduced in Refs.~\cite{Kirsanskas2018Jan,Mozgunov2020Feb,Nathan2020Apr}
and use them to investigate explicit models for the measurement apparatus,  Ref.~\cite{Steiner2020Apr} focuses on the stochastic nature of quantum measurements and provides an in-depth analysis of the measurement current.

\begin{acknowledgments}
We thank T. Karzig, F. Nathan, M. Rudner, J. Steiner, and F. von Oppen for discussions. This research was supported by the Danish National Research Foundation, the Danish Council for Independent Research $\vert$ Natural Sciences, and by the Microsoft Corporation. We also acknowledge funding by the Deutsche Forschungsgemeinschaft (DFG, German Research Foundation) under Grant No.~ 277101999, TRR 183 (project
C01 and Mercator program), under Germany's Excellence Strategy - Cluster of Excellence Matter
and Light for Quantum Computing (ML4Q) EXC 2004/1 - 390534769, and under Grant No.~EG 96/13-1.
\end{acknowledgments}

\appendix

\section{Bath correlator for QPC detector}\label{appA}

In this appendix, we derive the bath autocorrelator $\BQPC(\omega)$ in \Eq{eq_qpc_bath_correlation_fourier_final} for a quantum point contact capacitively coupled to the dot-MQB system, see Fig.~\ref{fig4}. We start from the interaction Hamiltonian \eq{eq_hamiltonian_qpc_interaction},
\begin{equation}\label{app:QPCHI}
	H_I = \frac{2e^2}{C_{m}} \hat\rho \, n_d,
\end{equation}
where $\hat\rho = \psi^\dagger \psi^{\phantom{\dagger}}$ is the electron density operator in a small (approximately point-like) volume $V$ centered around the longitudinal coordinate
$x=0$ along the QPC.  Near this point, the capacitive
coupling between the QPC charge density and the dot charge will be most pronounced.  Here, $\psi$ is the electron annihilation operator for QPC electrons in this volume, and $C_m$ is the mutual dot-QPC capacitance per volume. The electron spin is accounted for by the factor $2$ in Eq.~\eqref{app:QPCHI}.

As concrete example, we model the QPC as 1D fermion system connected to electron reservoirs on the left and right side, with chemical potentials $\mu_L$ and $\mu_R$, respectively.
We assume that the capacitive interaction involves the QPC charge density at $x=0$ only, see Eq.~\eqref{app:QPCHI}. The QPC itself is  modeled by a $\delta$-peak  barrier of height ${\cal V}_0$ per unit length.  The corresponding contribution to the first-quantized Hamiltonian is
$V_{\rm QPC} = {\cal V}_0 \delta(x)$. We next express the local QPC fermion operator $\psi$ as
\begin{equation}
	\psi = \sum_{\ell=L/R,k} \Psi_{\ell k}(x=0)\, c_{\ell k} ,
	\label{eq_psi_op}
\end{equation}
where the $c_{\ell k}$ are fermionic annihilation operators corresponding to the single-particle QPC scattering states $\Psi_{\ell k}(x)$ with wave number $k$ originating from reservoir $\ell=L,R$.
For the 1D QPC model with a $\delta$-barrier, one finds   \cite{Griffiths2018Aug}
\begin{align}
	&\Psi_{\ell=L/R,k}(x) = \frac{1}{\sqrt{L_0}}\Big[\left( e^{\pm i k x}  + r_k e^{\mp i k x} \right)\Theta(\mp x) \nonumber\\
	&\qquad\qquad\qquad \qquad+ t_k e^{\pm i k x} \Theta(\pm x) \Big],\label{app:QPCtransref}\\
	& \quad r_k = \frac{1}{i \frac{k}{m{\cal V}_0}-1 }, \quad t_k = \frac{1}{1 + i \frac{m{\cal V}_0}{k}},\nonumber
\end{align}
where $L_0$ is the QPC length, $m$ the electron mass, and $r_k$ and $t_k$ are reflection and transmission amplitudes, respectively.
The charge density at $x=0$ follows as
\begin{align}
	\hat\rho = \frac{1}{V} \sum_{\ell,\ell'=L,R}\sum_{kk'} &\tau^{}_{\ell k,\ell' k'}\, c^\dagger_{\ell k}c^{}_{\ell' k'},
\end{align}
where $\tau^{}_{\ell k,\ell'k'}$ quantifies the overlap between $\Psi_{\ell k}$ and $\Psi_{\ell' k'}$.
For the 1D model with Eq.~\eqref{app:QPCtransref}, we obtain
\begin{equation}
	\tau_{\ell k,\ell'k'}^{} = \frac{1}{4}(1+r_k + t_k)(1+r_{k'} + t_{k'}),
\end{equation}
which is independent of the lead indices $\ell,\ell'$.
For calculating the bath correlation function, we next assume
\begin{equation}\label{app:QCPavcc}
	\av{c_{\ell k}^\dagger c^{\phantom{\dagger}}_{\ell'k'}} = \delta_{\ell\ell'} \delta_{kk'} \nFa(\epsilon_{\ell k}),
\end{equation}
with Fermi-Dirac distribution functions, $\nFa(\epsilon) = 1/(e^{\beta(\epsilon - \mu_\ell)}+1)$, and the single-particle eigenenergies, $\epsilon_{\ell k}$, in the bath Hamiltonian $H_B$, see
\Eq{eq_hamiltonian_qpc_interaction}.  Equation~\eqref{app:QCPavcc} effectively enforces the constraint that electrons thermalize before entering the QPC.
We then have
\begin{equation}
	\av{\hat\rho(t)} = \av{\hat\rho} = \frac{1}{V} \sum_{\ell, k} \tau^{}_{\ell k,\ell k}\, \nFa(\epsilon_{\ell k}),\label{eq_density_linear_moment}
\end{equation}
where $\hat\rho(t) = e^{i H_B t}\hat\rho e^{-iH_B t}$ and $\tau^{}_{\ell k,\ell k} > 0$. For the 1D example with a $\delta$-barrier, we have $\tau_{\ell k,\ell k} = \abs{t_k}^2$ according to Eq.~\eqref{app:QPCtransref}.
With the bath operator $\varphi$ in \Eq{eq_hamiltonian_qpc_interaction}, we now observe that Eq.~\eq{eq_density_linear_moment} implies a time-independent linear moment,
\begin{equation}
	\av{\varphi(t)} = \av{\varphi} = E_{\text{ref}}\sum_{\ell,k} \tau^{}_{\ell k,\ell k} \nFa(\epsilon_{\ell k}).\label{eq_bath_linear_moment}
\end{equation}
Rewriting the interaction Hamiltonian as
\begin{equation}
	H_I =  \sqrt{g} n_d (\varphi - \av{\varphi})  + \sqrt{g}n_d\av{\varphi},
\end{equation}
we observe that the linear moment in Eq.~\eq{eq_bath_linear_moment} can be absorbed by a shift of the dot level energy $\epsilon$,
\begin{equation}
 H_I \rightarrow \sqrt{g}n_d(\varphi - \av{\varphi}),\quad \epsilon \rightarrow \epsilon - \sqrt{g}\av{\varphi},
 \end{equation}
see Eqs.~\eq{Hamil0} and \eq{eq_hamiltonian_qpc_interaction}.
With respect to the redefined interaction Hamiltonian, the time-dependent bath autocorrelator in Eq.~\eqref{autocor}, $B_{\rm QPC}(t)$, which
enters the effective Lindbladian approximation,  can be evaluated by using $c_{\ell k}(t) = e^{-i\epsilon_{\ell k} t} c_{\ell k}$ along with Wick's theorem and Eq.~\eqref{app:QCPavcc}.
The result is
\begin{align}\nonumber
	B_{\rm QPC}(t)& = E_{\text{ref}}^2 \sum_{\ell k,\ell' k'}\abs{\tau^{}_{\ell k,\ell'k'}}^2 e^{i(\epsilon_{\ell k} - \epsilon_{\ell'k'})t}\\
	\label{eq_qpc_bath_correlation_time}
	&\times \nFa (\epsilon_{\ell k})\left[1-\nFap(\epsilon_{\ell' k'}) \right].
\end{align}
We now introduce the coupling profile function $\Gamma_{\ell\ell'}(\omega,\omega')$ as in \Eq{eq_qpc_coupling}, which for the 1D case with a $\delta$-barrier is given by
\begin{align}
	\Gamma_{\ell\ell'}(\omega,\omega') =
	 \frac{\frac{m}{4\pi^2}L_0^2E_{\text{ref}}^2\sqrt{{\omega}{\omega'}}}{\left(\frac{m{\cal V}_0^2}{2}+{\omega}\right)\left(\frac{m{\cal V}_0^2}{2}+{\omega'}\right)}\Theta(\omega)\Theta(\omega'), \label{eq_coupling_1d}
\end{align}
where we use Eq.~\eqref{app:QPCtransref} and $\frac{1}{L_0}\sum_k (\cdots) \rightarrow \frac{1}{2\pi} \int dk (\cdots)$.
Identifying the general form \eq{eq_qpc_coupling} of $\Gamma$ in \Eq{eq_qpc_bath_correlation_time}, we find
\begin{align}
	&B_{\rm QPC}(t) = \sum_{\ell,\ell'} \int_{-\infty}^\infty d\omega d\omega'\; \Gamma(\omega,\omega')e^{i(\omega-\omega')t}\nonumber\\
	&\times   \nB(\omega-\omega'-\mu_\ell + \mu_{\ell'})\left [\nFap(\omega')- \nFa(\omega) \right],\label{eq_qpc_final_step}
\end{align}
where we used the identity
\[
	\nF(\xi)\left[1-\nF(\xi')\right] =  \nB(\xi-\xi')\left[\nF(\xi') - \nF(\xi)\right].
\]
 Changing variables in \Eq{eq_qpc_final_step} to $\Omega = (\omega+\omega')/2$ and $\nu=\omega-\omega'$,  shifting $\Omega$ by $\mu_{\ell\ell'} = (\mu_{\ell} + \mu_{\ell'})/2$, and finally performing a Fourier transformation, we arrive at  \Eqs{eq_qpc_bath_correlation_fourier} and \eq{eq_qpc_spectral_density}.

 Finally, we note that if we evaluate $\Gamma_{\ell\ell'}\left(\Omega + \mu_{\ell\ell'} - \frac{\omega}{2},\Omega + \mu_{\ell\ell'} + \frac{\omega}{2}\right)$ with the coupling function \Eq{eq_coupling_1d} for $E_{\text{ref}} = \mu_0/2$ with $\mu_0=(\mu_L+\mu_R)/2$ and $m{\cal V}_0^2 \ll \mu_0$ as well as $1/(mL_0^2)  \ll \mu_0$, we can qualitatively confirm the behavior of $\Gamma_{\ell\ell'}\left(\Omega + \mu_{\ell\ell'} - \frac{\omega}{2},\Omega + \mu_{\ell\ell'} + \frac{\omega}{2}\right)$ assumed below \Eq{eq_qpc_coupling} in order to arrive at the simplification in \Eq{eq_qpc_coupling_cutoff}.	 For all involved integrals to converge, $\Gamma$ is here assumed to decay sufficiently fast at large frequencies due to the finite electronic bandwidth in the leads.

\section{Matrix form of the Liouvillian}\label{AppB}

In this appendix, we specify the full matrix form of the Liouvillian.
For a generic environmental correlation function, $B(\omega)$, using the energy eigenstates $|p,s\rangle$ of the combined dot-plus-coupled-MBS system in Eq.~\eqref{eq_different_energies} and the quantities $\Delta_s$ in Eq.~\eqref{RotHam}, the jump operator takes the general form
\begin{align}
L &= -\frac{\sqrt{g}}{2} \sum_{s=\pm1} \Bigg[ \frac{\epsilon\sqrt{B(0)}}{E_s} ( \outprod{-,s}{-,s}- \outprod{+,s}{+,s})\notag\\
&\qquad\qquad \notag + \frac{ \Delta_s\sqrt{B(-E_s) }}{E_s} \outprod{+,s}{-,s}\\&\qquad \qquad +\frac{\Delta_s\sqrt{  B(E_s) }}{E_s} \outprod{-,s}{+,s}\Bigg]\label{JumpOpGenericB}.
\end{align}
Using the basis in Eq.~\eqref{BlockDiagBasis},  the matrix form of the Liouvillian contains the blocks $\mathcal{L}_i$ with $i=\pm, c$,
\begin{align}
\mathcal{L} &= \begin{pmatrix} \mathcal{L}_+ &0&0&0\\ 0& \mathcal{L}_c&0&0 \\ 0&0&\mathcal{L}_c^* &0 \\ 0&0&0&\mathcal{L}_-\end{pmatrix},
\end{align}
with the parity-diagonal blocks ($s=\pm 1$),
\begin{align}
\mathcal{L}_{s} &= \begin{pmatrix}-g A_s^+ &-g m_s&-g m_s& g A_s^- \\-g n_{s}^+ & -g B_s +iE_s & g C_s &g n_{s}^-\\ -g n_{s}^+ & g C_s & -g B_s - iE_s &g n_{s}^- \\ g A_s^+ &g m_s&g m_s& -g A_s^-\end{pmatrix},
\end{align}
and the matrix
\begin{align}
 \mathcal{L}_c = \begin{pmatrix}\mathcal{L}^{++}_c & \mathcal{L}^{+-}_c \\ \mathcal{L}^{-+}_c & \mathcal{L}^{--}_c \end{pmatrix}.
\end{align}
This matrix contains the following $2\times 2$ blocks:
\[
  \mathcal{L}^{+-}_c = \begin{pmatrix} g j_+^- & g \sqrt{A_+^- A_-^-} \\ g \sqrt{A_-^+ A_+^-} & g q_+^- \end{pmatrix},
\]
\[
  \mathcal{L}^{-+}_c = \begin{pmatrix} gq_+^+ & g\sqrt{A_-^- A_+^+} \\ g\sqrt{A_+^+ A_-^+}&g j_+^+ \end{pmatrix},
  \]
  \[
 \mathcal{L}^{++}_c =-\begin{pmatrix} \scriptstyle \frac{g}{2}\big(A_+^+ + A_-^+ \big) + g K^- - i f^- & \scriptstyle -g j_-^- \\ \scriptstyle -g q_-^+& \scriptstyle \frac{g}{2} \big(A_-^- + A_+^+ \big) + g K^+ - if^+ \end{pmatrix},
 \]
 \[
 \mathcal{L}^{--}_c = -\begin{pmatrix} \scriptstyle \frac{g}{2} \big(A_+^- + A_-^+ \big) + g K^+  + if^+ & \scriptstyle -g q_-^- \\ \scriptstyle -g j_-^+& \scriptstyle \frac{g}{2}\big(A_+^- + A_-^- \big) + g K^- + if^-\end{pmatrix}.
\]
In the above expressions, we have used the quantities $A_s^\pm$ and $n_s^\pm$ in Eq.~\eqref{DefnAsp}, and $K^\pm$ and $f^\pm$ in Eq.~\eqref{kfdef}. Moreover, we define
\begin{align}\notag
B_s &= \frac{1}{8E_s^2} \left[ \Delta_s^2 \left(B(E_s) + B(-E_s) \right) + 4\epsilon^2 B(0) \right],\\
C_s &= \frac{\Delta_s^2}{4 E_s^2} \sqrt{B(E_s)B(-E_s)},\\ \notag
m_s&= \frac{\Delta_s \epsilon}{8 E_s^2 }   \sqrt{B(0)}\left(\sqrt{B(E_s)}+\sqrt{B(-E_s)} \right),
\end{align}
as well as
\begin{align}
j_s^\pm &= \pm\frac{\Delta_s \epsilon}{8E_s^2 E_{-s}}  \sqrt{B(0)} \left[ \sqrt{B(\pm E_s)}(2 E_s - E_{-s})\right.\notag\\
&\phantom{= \pm\frac{\Delta_s \epsilon}{8E_s^2 E_{-s}}  \sqrt{B(0)} \left[\right.}\left.+ E_{-s} \sqrt{B(\mp E_s)}  \right],\\
q_s^\pm &= \pm \frac{\Delta_s \epsilon}{8E_s^2 E_{-s}}  \sqrt{B(0)} \left[ -\sqrt{B(\pm E_s)}(2 E_s + E_{-s})\right.\notag\\ \notag
&\phantom{= \pm \frac{\Delta_s \epsilon}{8E_s^2 E_{-s}}  \sqrt{B(0)} \left[\right.}\left. +E_{-s} \sqrt{B(\mp E_s)}  \right].
\end{align}
\normalsize

%

\end{document}